\documentclass[showpacs,preprint,aip,eqsecnum,amssymb,amsfonts,verbatim]{revtex4}

\usepackage{amsthm,graphicx}

\newcommand{\rmi}{{\rm i}}
\newcommand{\rme}{{\rm e}}
\newcommand{\rmd}{{\rm d}}

\begin{document}

\def\llra{\relbar\joinrel\longrightarrow}              
\def\mapright#1{\smash{\mathop{\llra}\limits_{#1}}}    
\def\mapup#1{\smash{\mathop{\llra}\limits^{#1}}}     
\def\mapupdown#1#2{\smash{\mathop{\llra}\limits^{#1}_{#2}}} 

\catcode`\@=11

\def\BF#1{{\bf {#1}}}
\def\NEG#1{{\rlap/#1}}

\def\Let@{\relax\iffalse{\fi\let\\=\cr\iffalse}\fi}
\def\vspace@{\def\vspace##1{\crcr\noalign{\vskip##1\relax}}}
\def\multilimits@{\bgroup\vspace@\Let@
 \baselineskip\fontdimen10 \scriptfont\tw@
 \advance\baselineskip\fontdimen12 \scriptfont\tw@
 \lineskip\thr@@\fontdimen8 \scriptfont\thr@@
 \lineskiplimit\lineskip
 \vbox\bgroup\ialign\bgroup\hfil$\m@Tu\scriptstyle{##}$\hfil\crcr}
\def\Sb{_\multilimits@}
\def\endSb{\crcr\egroup\egroup\egroup}
\def\Sp{^\multilimits@}
\let\endSp\endSb


\title[The RHS approach to the Gamow states]{The rigged Hilbert space approach 
to the Gamow states}

\author{Rafael de la Madrid}
\affiliation{Department of Physics, Lamar University,
Beaumont, TX 77710 \\
E-mail: \texttt{rafael.delamadrid@lamar.edu}}


\date{\today}

\begin{abstract}
\noindent We use the resonances of the 
spherical shell potential to present a thorough description of the Gamow
(quasinormal) states within the rigged Hilbert space. It will be 
concluded that the natural setting for the Gamow states is a 
rigged Hilbert space whose test functions fall off at infinity
faster than Gaussians.
\end{abstract}


\maketitle

\section{Introduction}
\label{sec:introduction}

Resonances are intrinsic properties of a quantum system, and they describe 
the system's preferred ways of decaying. The experimental fingerprints 
of a resonance are either a sharp peak in the 
cross section or the exponential decay of the probability to find the
unstable particle. The sharp peaks in the cross section are
characterized by the energy $E_{\rm R}$ at which they occur and by their
width $\Gamma _{\rm R}$. Decay is characterized by the energy $E_{\rm R}$ 
of the particle and by its lifetime $\tau _{\rm R}$.

The Gamow states are the wave functions of resonances, and they
are eigenvectors of the Hamiltonian with a complex 
eigenvalue. The real part of the complex eigenvalue is associated with 
the energy of the resonance, and the imaginary part is associated with 
the width. The time evolution of the Gamow eigenfunctions abides by 
the exponential decay law. 

The Gamow states are able to describe both sharp peaks in the cross section 
and decay, in accordance with the phenomenological perception that resonances 
and unstable particles are two sides of the same phenomenon. As well, when
the (complex) resonance energy tends to a (real) bound-state energy, the
Gamow eigenfunction becomes a bound state, in accordance with the
phenomenological perception that unstable states are only quantitatively, 
not qualitatively, different from bound states, the only difference being 
that unstable states have a non-zero width, whereas the width of stable 
states is zero.

In a way, the Gamow states complete the so-called Heisenberg program,
according to which spectral lines, widths and lifetimes are all observable 
quantities, and quantum mechanics should be able to predict them. 

Gamow introduced the energy eigenfunction with complex eigenvalue in 
his paper on $\alpha$-decay of
atomic nuclei~\cite{GAMOW}, and its properties and 
applications have been considered by many authors, see for 
example~\cite{SIEGERT,PEIERLS55,PEIERLS59,HUMBLET,ZELDOVICH,BERGGREN,
MORE,ROMO,gcp76,WOLF,BERGGREN78,SUDARSHAN,MONDRAGON84,
FERREIRA1,BGM,CURUTCHET,KUKULIN,
BG,BL,LIND,VERTSE,BOLLINI1,BOLLINI2,BERGGREN96,FERREIRA2,GADELLA,
TOLSTIKHIN,TOLSTIKHIN2,CAVALCANTI,MONDRADOUBLE,FERREIRA3,BETAN,
MICHEL02,AJP02,KAPUSCIK1,MICHEL2,
MONDRAGON03,CAVALCANTI03,KAPUSCIK,CIVITARESE,MICHEL05,SANTRA,
05CJP,DELION,MONDRA06,MICHEL3,MICHEL4,STRAUSS06,
GASTON07,MICHEL5,MICHEL6,MONDRA07a,MONDRA07b,COCOYOC,
MICHEL-SOLO,URRIES,COSTIN,
VELAZQUEZ,MICHEL08,
ROSAS22,TOMIO,NPA08,
MICHEL7,HATANO,HUANG,GRUMMT09,
HATANO10,GASTON10,GOUSEEV10,JULVE10,
MICHEL10,VECCHI11,
ROSAS11,STRAUSS11a,STRAUSS11b,ROSASS11b,COSTIN11,DURR11,POLACO,GASTON11,
DELCAMPO11,RAPEDIUS11,HATANO11,BELCHEV11,FERNANDEZ11}
and references therein. A pedestrian introduction 
to these states can be found in~\cite{BGM,AJP02,CAVALCANTI}. Gamow's treatment 
does not fit within the Hilbert space though, because 
self-adjoint operators on a Hilbert space can only have real 
eigenvalues. Recall however that Dirac's bra-ket formalism does not fit
within the Hilbert space but rather within the rigged Hilbert 
space. Similarly, the rigged Hilbert space 
mathematics asserts the legitimacy of Gamow's proposition. In the rigged
Hilbert space language, the Gamow states are 
eigenvectors of the dual extension of the self-adjoint Hamiltonian. Such 
extension can surely have complex eigenvalues~\cite{LINDBLAD}.

A rigged Hilbert space (also called a Gelfand triplet) is a triad of spaces
\begin{equation}
       {\mathbf \Phi} \subset {\cal H} \subset {\mathbf \Phi}^{\times}
              \label{RHS}
\end{equation}
such that ${\cal H}$ is a Hilbert space, $\mathbf{\Phi}$ is a dense subspace 
of~${\cal H}$, and ${\mathbf \Phi}^{\times}$ is the anti-dual space
of $\mathbf \Phi$. The space $\mathbf \Phi$ has a topology that is finer than 
the topology inherited from $\cal H$. The space ${\mathbf \Phi}^{\times}$ 
contains the continuous, antilinear functionals 
over $\mathbf \Phi$. Associated with the rigged Hilbert 
space~(\ref{RHS}), there is always another rigged Hilbert space,
\begin{equation}
       {\mathbf \Phi} \subset {\cal H} \subset {\mathbf \Phi}^{\prime} \, , 
              \label{RHS-du}
\end{equation}
where ${\mathbf \Phi}^{\prime}$ is called the dual space of $\mathbf{\Phi}$ and 
contains the continuous, linear functionals 
over~$\mathbf \Phi$. Since the space ${\mathbf \Phi}^{\times}$ is bigger 
than ${\cal H}^{\times}\equiv {\cal H}$, and since ${\mathbf \Phi}^{\prime}$ 
is also bigger than ${\cal H}^{\prime}\equiv {\cal H}$, some physically 
meaningful states that find no accommodation in $\cal H$ will find 
accommodation in ${\mathbf \Phi}^{\times}$ and ${\mathbf \Phi}^{\prime}$. For 
example, the eigensolutions of the
time-independent Schr\"odinger equation associated with either the
scattering energies (the Lippmann-Schwinger kets $|E^\pm \rangle$) or 
with the resonant energies (the Gamow kets $|z_{\rm R}\rangle$) find 
accommodation in ${\mathbf \Phi}^{\times}$, whereas the bras 
$\langle ^{\pm}E|$ and $\langle z_{\rm R}|$ find accommodation in 
${\mathbf \Phi}^{\prime}$.


The present paper is devoted to show how the rigged Hilbert space 
is able to accommodate the Gamow states. Throughout the paper, rather 
than working in a general setting, we will use 
the example of the spherical shell potential,
\begin{equation}
          V({\bf x})= V(r)=\left\{ \begin{array}{ll}
                                0   &0<r<a  \\
                                V_0 &a<r<b  \\
                                0   &b<r<\infty \, ,
                  \end{array} 
                 \right. 
	\label{potential}
\end{equation}
and restrict ourselves to the s partial wave. However, as explained in 
Appendix A of Ref.~\cite{NPA08}, the result is valid for any partial wave 
and for spherically symmetric potentials that fall off
faster than exponentials.

For the potential~(\ref{potential}), expressions such as those for the Gamow 
eigenfunctions and the $S$ matrix depend on the square root of the energy
rather than on the energy itself. It is, therefore, easier to do calculations 
with the wave number $k$,
\begin{equation}
      k=\sqrt{\frac{2m}{\hbar ^2}E\,} \, , 
        \label{wavenumber}
\end{equation}      
rather than with the energy $E$. However, we will write most 
results in terms of the energy, because they tend to be simpler than in
terms of the wave number. Also, when the energy and the wave number become 
complex, we will denote them by $z$ and $q$,
\begin{equation}
      q=\sqrt{\frac{2m}{\hbar ^2}z\,} \, , 
        \label{wavenumberc}
\end{equation} 
and when they correspond to a resonance 
${\rm R}$, we will denote them by $z_{\rm R}$ and $k_{\rm R}$,
\begin{equation}
      k_{\rm R}=\sqrt{\frac{2m}{\hbar ^2}z_{\rm R} \,} \, . 
        \label{wavenumberR}
\end{equation} 
We will re-write most expressions in Dirac's bra-ket notation,
because of its simplicity, clarity and beauty. 

Wave functions in the position representation, denoted by $\varphi$, and
Gamow bras and kets, denoted by $\langle z_{\rm R}|$ and $|z_{\rm R}\rangle$,
will have sometimes a superscript $+$ or $-$ attached to them, and it is
important to understand what this superscript means. Let us suppose that 
$\varphi$ is a Gaussian wave packet in the position representation. Such
Gaussian could be either an ``in'' state, in which case we denote it by
$\varphi ^+$, or an ``out'' state, in which case we denote it by 
$\varphi ^-$. When we write $\varphi ^+$, the Gaussian will be expanded 
by the ``in'' Lippmann-Schwinger bras and kets, and its energy representation 
will always be the one associated with the ``in'' bras and kets. When 
we write $\varphi ^-$, the Gaussian will be expanded by the ``out'' 
Lippmann-Schwinger bras and kets, and its energy representation will always 
be the one associated with the ``out'' bras and kets. Thus, the 
superscripts~$\pm$ are sort of ``phase-space'' labels, since they 
tell us which energy 
representation we are using, even though we may be working in the
position representation. Physically, the superscripts~$\pm$ are a 
reminder of whether we have imposed the ``in'' or the ``out'' boundary 
conditions on the Gaussian packet. For the Gamow bras and kets, the 
meaning of the superscripts~$\pm$ is analogous. 

Of all the previous attempts to describe the Gamow states within
the rigged Hilbert space, our approach is closest to that of
Bollini {\it et al.}~\cite{BOLLINI1,BOLLINI2}. There are, however, two main 
differences between the present approach and that of 
Refs.~\cite{BOLLINI1,BOLLINI2}. First, the test functions we are
going to use fall off at infinity faster
than Gaussians, whereas the test functions used in 
Refs.~\cite{BOLLINI1,BOLLINI2} fall off
at infinity faster than exponentials. We use test 
functions that fall off faster than Gaussians because they enable us
to perform resonance expansions that include all the resonances of the
system and that exhibit time asymmetry~\cite{VANTONDER}.
And second, we obtain the
relation between the Breit-Wigner amplitude and the Gamow states by 
transforming to the energy representation, whereas in 
Refs.~\cite{BOLLINI1,BOLLINI2} such relation is obtained by transforming
to the momentum representation.


In Sec.~\ref{sec:gamowvectors}, the Gamow states of the spherical shell 
potential will be constructed. The Gamow kets associated with
resonances and anti-resonances will be defined as the 
solutions of homogeneous integral equations of the Lippmann-Schwinger 
type. We will solve these integral equations in the radial, position 
representation. In this representation, those integral 
equations are equivalent to the time-independent Schr\"odinger equation 
subject to a purely outgoing boundary condition (POBC). We will also obtain 
the ``left'' Gamow eigenfunctions and will
comment on the analogy between bound and resonance states.

In Sec.~\ref{sec:GSdis}, we will apply the theory of distributions
to construct the Gamow bras and kets, which in Sec.~\ref{sec:GveRHS} will be 
shown to be generalized eigenvectors of the Hamiltonian with 
complex eigenvalues. Also in Sec.~\ref{sec:GveRHS}, we will construct
the rigged Hilbert spaces that accommodate the Gamow bras and kets.

Next, in Sec.~\ref{sec:Gvecenwnrepr}, we will obtain the energy 
representations of the Gamow bras and kets, and show that they can be
written in terms of the complex delta function and the residue distribution.

In Sec.~\ref{sec:theminpinerep}, we will let the energy run over the
full real line in order to obtain the ``energy representation'' associated
with the Breit-Wigner distribution. We will show how the complex delta
function becomes the Breit-Wigner distribution in such ``energy 
representation.'' The results of Secs.~\ref{sec:Gvecenwnrepr} 
and~\ref{sec:theminpinerep} will, in particular, provide a mathematical
support for the results presented in Ref.~\cite{NPA08}.

The time evolution of the Gamow bras and kets will be calculated in
Sec.~\ref{sec:semievolut}. We will argue, although not fully prove, 
that the 
time evolution of a resonance ket is valid for positive times only, 
whereas the time evolution of an anti-resonance ket is valid for negative 
times only. Thus, the time evolution of resonances is given by (non-unitary)
semigroups, which express the time asymmetry built into a decaying
process. This time asymmetry seems to be what some authors such as 
Fonda {\it et al.}~\cite{FONDA}, Cohen-Tannoudji {\it et al.}~\cite{COHEN}, 
or Goldberger and Watson~\cite{GOLDBERGER} have called 
the {\it irreversibility} of a decaying process.

For the sake of completeness, in Sec.~\ref{sec:resexp} we will construct
the resonant expansions and see how such expansions allow us
to isolate each resonance's contribution and to interpret the 
deviations from exponential decay~\cite{RAIZEN}.


In Sec.~\ref{sec:phmean}, we will present two analogies that help to understand
the physical meaning of the Gamow states. The first analogy is that between the
resonance expansions, the Dirac expansions, and the classical Fourier 
expansions. The second analogy is that between the classical,
quasinormal modes and the quantum mechanical resonances. We will also
explain the physical reason why the Gamow eigenfunctions blow up exponentially 
at infinity.




\section{The Gamow eigenfunctions}
\label{sec:gamowvectors}

The Gamow eigenfunctions are customarily defined as eigensolutions of the 
Schr\"odinger equation subject to the POBC. Although we 
could start the study of the Gamow states with that definition, we will follow 
instead a treatment parallel to that of the Lippmann-Schwinger
equation~\cite{DIS,LS1,LS2}. We will define a Gamow state as the solution of an 
integral equation~\cite{WOLF,MONDRAGON84} that has the POBC built 
into it. Needless to say, in the end the explicit solutions of that integral 
equation will be found by solving the Schr\"odinger equation subject to the 
POBC.

\subsection{The integral equation of the Gamow states}
\label{sec:LSEofGV}

The Gamow states are solutions of a homogeneous integral equation of the 
Lippmann-Schwinger type. If $z_{\rm R}=E_{\rm R}- {\rm i} \Gamma _{\rm R} /2$ 
denotes the complex energy associated with a resonance of energy $E_{\rm R}$ 
and width $\Gamma _{\rm R}$, then the corresponding Gamow state 
$|z_{\rm R}\rangle$ fulfills~\cite{WOLF,MONDRAGON84}
\begin{equation}
    |z_{\rm R}\rangle =\frac{1}{z_{\rm R}-H_0+\rmi 0}V|z_{\rm R}\rangle  \, .
       \label{Monlisus}
\end{equation}
The $+\rmi 0$ in Eq.~(\ref{Monlisus}) means that we are 
working with the retarded free Green function, which has a purely outgoing 
boundary condition built into it. The retarded free Green function is 
analytically continued across the cut into the lower half plane of the 
second sheet of the Riemann surface, where the complex number $z_{\rm R}$ is 
located. Therefore, as pointed out in~\cite{MONDRAGON84}, 
Eq.~(\ref{Monlisus}) should be written as
\begin{equation} 
       |z_{\rm R}\rangle =\lim_{E\to z_{\rm R}}
                 \frac{1}{E-H_0+\rmi 0}V|E\rangle \, .
       \label{goodMonlisus}
\end{equation}
This notation intends to express that we first have
to calculate the retarded free Green function $(E-H_0+\rmi 0)^{-1}$ in the
physical sheet, and then continue it across the cut into the lower half plane
of the second sheet. 

The integral equation~(\ref{Monlisus}) has the POBC built 
into it. To be more precise, in the position representation 
Eq.~(\ref{Monlisus}) is equivalent to the time-independent Schr\"odinger 
equation subject to the condition that far away from the 
potential region, the solution behave as a purely outgoing wave.

As is well known, to each resonance energy $z_{\rm R}$ there corresponds
an anti-resonance energy $z_{\rm R}^*$ that lies in the upper half plane
of the second sheet. The integral equation satisfied by the anti-resonance 
state $|z_{\rm R}^*\rangle$ reads as
\begin{equation}
       |z_{\rm R}^*\rangle =
             \frac{1}{z_{\rm R}^*-H_0-\rmi 0}V|z_{\rm R}^*\rangle =
       \lim_{E\to z_{\rm R}^*}\frac{1}{E-H_0-\rmi 0}V|E\rangle \, .
       \label{groMonlisus}
\end{equation}
In contrast to Eq.~(\ref{Monlisus}), Eq.~(\ref{groMonlisus}) has a purely
{\it incoming} boundary condition built into it. That is, 
in the position representation, Eq.~(\ref{groMonlisus}) is equivalent to the 
time-independent Schr\"odinger equation subject to the condition that far 
away from the potential region, the solution behave as a purely incoming wave.

\subsection{The Gamow states in the position representation}
\label{sec:Gsavepsire}

In the radial, position representation, Eqs.~(\ref{Monlisus}) and 
(\ref{groMonlisus}) become
\begin{eqnarray}
       \langle r|z_{\rm R}\rangle =
         \langle r|\frac{1}{z_{\rm R}-H_0+\rmi 0}V|z_{\rm R}\rangle
       =\lim_{E\to z_{\rm R}} 
           \langle r|\frac{1}{E-H_0+\rmi 0}V|E\rangle \, ,
           \label{posGinte1} \\
       \langle r|{z_{\rm R}^*}\rangle =
       \langle r|\frac{1}{z_{\rm R}^*-H_0-\rmi 0}V|{z_{\rm R}^*}\rangle =
       \lim_{E\to z_{\rm R}^*}\langle r|\frac{1}{E-H_0-\rmi 0}V|E\rangle \, .
            \label{posGinte2}
\end{eqnarray}
In~\cite{MONDRAGON84}, these integral equations are written as
\begin{equation}
   u(r;z_{\rm R}) =\lim _{E\to z_{\rm R}}\int_0^{\infty}
         G_0^+(r,s;E)V(s)u(s;E) \, \rmd s \, , 
            \label{integrenot}
\end{equation}
\begin{equation}
   u(r;z_{\rm R}^*) =\lim _{E\to z_{\rm R}^*}\int_0^{\infty}
                G_0^-(r,s;E)V(s)u(s;E) \, \rmd s \, ,
          \label{integanrenot}
\end{equation}
where
\begin{equation}
       u(r;z_{\rm R})= \langle r|z_{\rm R}\rangle \, .
\end{equation}

In order to obtain the explicit expressions of the
Gamow eigenfunctions, instead of solving the integral 
equations~(\ref{integrenot}) and (\ref{integanrenot}), we solve the 
equivalent Schr\"odinger differential equation
\begin{equation}
       \left( -\frac{\hbar ^2}{2m}\frac{\rmd ^2}{\rmd r^2}+V(r)\right) 
       u(r;z_{\rm R}) = z_{\rm R} \, u(r;z_{\rm R}) \, ,
	\label{Grse0}
\end{equation}
subject to the boundary conditions built into those integral equations,
\begin{eqnarray}
	&& u(0;z_{\rm R}) = 0 \, , \label{gvlov1}  \\
	&& u(r;z_{\rm R}) \ \mbox{is continuous at} \ r=a,b \, ,  
                        \label{gvlov2}   \\
	&&\frac{\rmd}{\rmd r}u(r;z_{\rm R}) \ 
                     \mbox{is continuous at} 
          \ r=a,b \, ,  \label{gvlov5} \\
        && u(r;z_{\rm R}) \sim \rme ^{\rmi k_{\rm R}r} \ 
                       \mbox{as} \  r\to \infty \, , 
          \label{gvlov6}  
\end{eqnarray}
where condition~(\ref{gvlov6}) is the POBC. In 
Eqs.~(\ref{Grse0})-(\ref{gvlov6}), 
$u(r;z_{\rm R})\equiv \langle r|z_{\rm R} \rangle$ can denote either
a resonance or an anti-resonance state. 


For the spherical shell potential~(\ref{potential}), the only possible
eigenvalues of Eq.~(\ref{Grse0}) subject to~(\ref{gvlov1})--(\ref{gvlov6}) are
the solutions of the following transcendental equation:
\begin{equation}
      {\cal J}_+(z_{\rm R})=0 \, ,
         \label{ressoncon}
\end{equation}
where ${\cal J}_+$ is the Jost function, see, e.g., Refs.~\cite{DIS,AJP02}. The 
solutions of this equation come as a denumerable number of complex 
conjugate pairs $z_n, z_n^*$. The number $z_n=E_n -\rmi \Gamma _n /2$ is the 
$n$th resonance energy. The number $z_n^*=E_n +\rmi \Gamma _n /2$ is the 
$n$th anti-resonance energy. 
The corresponding resonance and anti-resonance 
wave numbers are given by
\begin{equation}
      k_n=\sqrt{\frac{2m}{\hbar ^2}z_n\,} \, , \quad
      -k_n^*=\sqrt{\frac{2m}{\hbar ^2}z_n^*\,} \, , \quad n=1,2, \ldots \, , 
\end{equation}
which belong, respectively, to the fourth and third quadrants of the 
$k$-plane. For the potential~(\ref{potential}), the resonance poles are 
simple (see~\cite{MONDRADOUBLE} for an example of a potential that
produces double poles).

In terms of the wave number $k_n$, the $n$th Gamow eigensolution 
reads 
\begin{equation}
      u(r;z_n)=u(r;k_n)= N_n\left\{ \begin{array}{ll}
         \frac{1}{{\mathcal J}_3(k_n)}\sin(k_{n}r)  &0<r<a \\ [1ex]
         \frac{{\mathcal J}_1(k_n)}{{\mathcal J}_3(k_n)}\rme ^{\rmi Q_{n}r}
         +\frac{{\mathcal J}_2(k_n)}{{\mathcal J}_3(k_n)}
                \rme ^{-\rmi Q_{n}r} &a<r<b 
         \\  [1ex]
         \rme ^{\rmi k_{n}r}  &b<r<\infty \, ,
                           \end{array} 
                  \right.
	\label{dgv0p} 
\end{equation}
where
\begin{equation}
      Q_n=\sqrt{\frac{2m}{\hbar ^2}(z_n-V_0)\,} \, ,
\end{equation}
$N_n$ is a normalization factor,
\begin{equation}
       N_n^2= \rmi \, \mbox{res} \left[ S(q) \right]_{q=k_n} \, ,
\end{equation}
and ${\cal J}_1$--${\cal J}_3$ are coefficients whose expressions 
follow from the matching conditions~(\ref{gvlov2}) and (\ref{gvlov5}). The 
Gamow eigensolution associated with the $n$th anti-resonance pole reads
\begin{equation}
     \hskip-1cm   u(r;z_n^*)=u(r;-k_n^*) =M_n\left\{ \begin{array}{ll}
         \frac{1}{{\mathcal J}_3(-k_n^*)}\sin(-k_{n}^*r)  &0<r<a \\ [1ex]
         \frac{{\mathcal J}_1(-k_n^*)}{{\mathcal J}_3(-k_n^*)}
          \rme ^{-\rmi Q_{n}^*r}
         +\frac{{\mathcal J}_2(-k_n^*)}{{\mathcal J}_3(-k_n^*)}
           \rme ^{\rmi Q_{n}^*r} &a<r<b \\ [1ex]
            \rme ^{-\rmi k_{n}^*r}  &b<r<\infty \, ,
                           \end{array} 
                  \right.
	\label{ggv0p} 
\end{equation}
where $M_n$ is a normalization factor,
\begin{equation}
       M_n^2= \rmi \, \mbox{res} \left[ S(q) \right]_{q=-k_n^*}=(N_n^2)^*  \, ,
\end{equation}
and where
\begin{equation}
       -Q_n^*=\sqrt{\frac{2m}{\hbar ^2}(z_n^*-V_0)\,} \, .
\end{equation}
For the sake of brevity, we will label the anti-resonance wave numbers
$-k_n^*$ and $-Q_n^*$, the energies $z_n^*$, the normalization factors
$M_n$ and the eigenfunctions $u(r;z_n^*)$ with a negative integer $n$ as
\begin{equation}
         k_n \, , \  Q_n \, , \  z_n \, , \  N_n \, , \ u(r;z_n) \qquad 
          n=-1,-2,\ldots \, .
\end{equation}
This notation will enable us to write results that are true for both 
resonances and anti-resonances just once.

Since they are eigenfunctions of a linear differential operator, the Gamow 
eigenfunctions~(\ref{dgv0p}) and (\ref{ggv0p}) are defined up to a 
normalization factor. The normalization we have adopted was introduced by 
Zeldovich~\cite{ZELDOVICH}, who used a Gaussian regulator to damp the
exponential blowup of the Gamow eigenfunctions and obtain a meaningful
normalization:
\begin{equation}
           \lim_{\mu \to 0}
              \int_0^{\infty} \rmd r \, \rme ^{-\mu r^2}
             [u(r;z_n)]^2 = 1 \, ,
               \label{Zregula}
\end{equation}
where $n=\pm 1 , \pm 2 \ldots$. Zeldovich's normalization has 
(at least) three advantages. First,
it generalizes the normalization of bound states; second, the residue of 
the propagator at the resonance energy factors out as a product of 
two Gamow eigenfunctions, see Eq.~(\ref{residue1}) below; and third,
Zeldovich's normalization makes $u(r;z_n)$ have dimensions of 
$1/ \sqrt{\rm length}$, so $|u(r;z_n)|^2$ has dimensions of a radial
probability density, just like any normalized wave function in the position 
representation.

It is worthwhile noting that the expressions for the delta-normalized 
Lippmann-Schwinger eigenfunctions are different when expressed in terms of 
$k$ from when expressed in terms of $E$~\cite{LS1,LS2}. However, 
similarly to bound states, the expressions for the normalized Gamow 
eigenfunctions are the same when expressed in terms of $k_{\rm R}$ as when
expressed in terms of $z_{\rm R}$, see Eqs.~(\ref{dgv0p}) and (\ref{ggv0p}).

\subsection{The ``left'' Gamow eigenfunctions}

After having obtained the ``right'' Gamow eigenfunctions, which will be
associated with the Gamow kets, it is easy to obtain the ``left'' 
Gamow eigenfunctions, which will be associated with the Gamow bras.

The ``left'' Gamow eigenfunctions can be obtained by complex Hermitian
conjugation of the ``right'' Gamow eigenfunctions~\cite{CHC}, or by 
analytic continuation
of the ``left'' Lippmann-Schwinger eigenfunctions~\cite{LS2,SIGMA}. The
resulting ``left'' Gamow eigenfunction associated with the resonance 
(or anti-resonance) energy $z_n$ is given by
\begin{equation}
       \langle z_n|r\rangle = [u(r;z_n^*)]^* 
      \, , \quad n=\pm 1, \pm 2, \ldots \, .
               \label{legazn1}
\end{equation}
Thus, contrary to naive expectations, the ``left'' Gamow eigenfunction is not 
just the complex conjugate of the ``right'' eigenfunction, but the complex
conjugated eigenfunction evaluated at the complex conjugated 
energy. Note that this 
procedure to obtain the ``left'' from the ``right'' eigenfunctions
generalizes the procedure to obtain the ``left'' from the ``right'' 
eigenfunctions of both the bound and the scattering 
eigenfunctions.
Because the Gamow eigenfunctions satisfy
\begin{equation}
           [u(r;z_n^*)]^* = u(r;z_n) \, , \quad n=\pm 1, \pm 2, \ldots \, ,
               \label{legazn2}
\end{equation}
the ``left'' and the ``right'' Gamow eigenfunctions are actually the same 
eigenfunction,
\begin{equation}
       \langle z_n|r\rangle = [u(r;z_n^*)]^* = u(r;z_n) = 
         \langle r|z_n \rangle \, , \quad n=\pm 1, \pm 2, \ldots \, .
               \label{legazn}
\end{equation}
In terms of the wave number, Eq.~(\ref{legazn}) reads as
\begin{equation}
       \langle k_n|r\rangle = [u(r;-k_n^*)]^* = u(r;k_n) = 
         \langle r|k_n \rangle \, , \quad n=\pm 1, \pm 2, \ldots \, .
               \label{legakn}
\end{equation}
Note that Eq.~(\ref{legazn2}) is a 
symmetry of the Gamow eigenfunctions, and it is such symmetry what 
in the end makes the ``left'' eigenfunction be equal to the ``right'' 
one. Note also that such symmetry does in general not hold when we change the 
normalization of the Gamow eigenfunctions---yet another reason to choose 
Zeldovich's normalization.

Equation~(\ref{legazn}) makes it clear why Zeldovich's 
normalization for the Gamow states is written as in~(\ref{Zregula}) rather than
as
\begin{equation}
          \lim_{\mu \to 0}
              \int_0^{\infty} \rmd r \, \rme ^{-\mu r^2}
             |u(r;z_n)|^2 =1  \, .
               \label{Zregulawrong}
\end{equation}
Also, Eq.~(\ref{legazn}) can be used to show that at a resonance (or 
anti-resonance) pole, the residue of the Green function is given by
\begin{equation}
     {\rm res} \left[ G(r,s;z) \right]_{z=z_n} = \frac{\hbar ^2}{m}k_n \,
      {\rm res} \left[ G(r,s;q) \right]_{q=k_n} = u(r;k_n)\, u(s;k_n) \, , 
       \quad n=\pm 1, \pm 2,\ldots  ,
          \label{residue1}
\end{equation}
which in bra-ket notation becomes
\begin{equation}
       {\rm res}[\langle r|\frac{1}{z-H}|s\rangle ]_{z=z_n} = 
        \langle r|z_n\rangle \langle z_n|s\rangle \, ,
        \quad n=\pm 1, \pm 2,\ldots  \, .
\end{equation}
Note that this 
factorization could have been used to define the
above normalization of the Gamow states and to show that the ``left''
Gamow eigenfunction $\langle z_n|s\rangle$ is the same as the ``right'' 
Gamow eigenfunction $\langle s|z_n\rangle$.

\subsection{Bound states}

For the sake of simplicity in the expressions, we have chosen a potential that
doesn't bind bound states. We would nevertheless like to briefly comment 
on what happens when bound states appear. 

The bound states satisfy the same integral equation as the resonance states, 
and therefore they automatically follow from the Schr\"odinger equation 
subject to the POBC along with resonances. Thus, the eigenfunction 
$u (r;z_{\rm R})$ becomes a bound state when we substitute the complex 
resonance energy $z_{\rm R}$ by a real bound-state energy $E_{\rm B}$. In 
addition, Zeldovich's normalization for the Gamow eigenfunctions reduces 
to the standard normalization of bound states when we substitute 
$z_{\rm R}$ by $E_{\rm B}$.




\section{The Gamow bras and kets}
\label{sec:GSdis}

The Gamow eigenfunctions $u(r;z_n)$ are obviously not square integrable, 
i.e., they do not belong to the Hilbert space $L^2([0,\infty ),\rmd r)$. Thus,
like the Lippmann-Schwinger eigenfunctions~\cite{LS1,LS2,DIS}, the
Gamow eigenfunctions must be treated as distributions. By treating them as
distributions, we will be able to generate the Gamow bras and kets.

According to the theory of distributions~\cite{GELFANDIII},
the Gamow ket $|z_n\rangle$ associated with the eigenfunction $u(r;z_n)$ must
be defined as~\cite{LS1,LS2,DIS}
\begin{equation}
     \hskip-1cm
   \begin{array}{rcl}
      |z_n\rangle : {\mathbf \Phi}_{\rm exp} & \longmapsto & {\mathbb C} 
        \nonumber \\ 
       \varphi & \longmapsto & \langle \varphi |z_n\rangle :=
        \int_0^{\infty} \rmd r\, [\varphi (r)]^* \,  u(r;z_n) 
        \, , \quad  n=\pm 1, \pm 2, \ldots \, .
             \label{Gketdef}
    \end{array}
\end{equation}
The elements $\varphi (r)$ of ${\mathbf \Phi}_{\rm exp}$ are such that
their ``nice behavior'' compensates the ``bad behavior'' of $u(r;z_n)$ so the 
integral~(\ref{Gketdef}) makes sense. The space ${\mathbf \Phi}_{\rm exp}$
will be constructed in Sec.~\ref{sec:GveRHS}. In the bra-ket notation, 
definition~(\ref{Gketdef}) becomes
\begin{equation}
    \langle \varphi |z_n\rangle =
        \int_0^{\infty} \rmd r\, \langle \varphi |r\rangle
           \langle r|z_n\rangle \, .
             \label{Gketdefb-k}
\end{equation}

Similarly, the Gamow bras associated with the resonance (or 
anti-resonance) energy $z_n$ are defined as
\begin{equation}
          \hskip-1cm
     \begin{array}{rcl}
      \langle z_n| : {\mathbf \Phi}_{\rm exp} & \longmapsto & {\mathbb C} 
        \nonumber \\ 
       \varphi  & \longmapsto & \langle z_n| \varphi \rangle :=
        \int_0^{\infty} \rmd r\, \varphi (r) u(r;z_n)  \, , \quad
            n= \pm 1, \pm 2 , \ldots  \, ;
             \label{Gbradef}
     \end{array}
\end{equation}
that is,
\begin{equation}
    \langle z_n| \varphi  \rangle =
        \int_0^{\infty} \rmd r\, 
                   \langle z_n|r\rangle \langle r|\varphi \rangle \, .
             \label{Gbradefb-k}
\end{equation}

From the above definitions and from Eq.~(\ref{legazn}),
it follows that the actions of the Gamow bras and kets are related by
\begin{equation}
    \langle \varphi |z_n \rangle = 
          \langle z_n^*|\varphi \rangle ^* \, , \quad 
        n= \pm 1, \pm 2 , \ldots \, .
\end{equation}

Since the Gamow eigenfunctions are the same when expressed in terms of the 
energy as when expressed in terms of the wave number, the Gamow bras and 
kets, unlike the delta-normalized Lippmann-Schwinger bras and kets, are 
the same when expressed in terms of the energy as when expressed in terms 
of the wave number:
\begin{equation}
       \hskip-0.5cm   
    |k_n\rangle = |z_n \rangle \, ,
       \qquad 
    \langle k_n| = \langle z_n| \, , 
       \quad n= \pm 1, \pm 2, \ldots  \, .
\end{equation}

\section{The rigged Hilbert spaces for the Gamow bras and kets}
\label{sec:GveRHS}

Likewise any bra or ket, the Gamow bras and kets are dealt with by means of 
the rigged Hilbert space rather than just by the Hilbert space. The 
rigged Hilbert space we will use is very similar to, although not the 
same as the rigged Hilbert space of Refs.~\cite{BOLLINI1,BOLLINI2}. We 
will denote the rigged Hilbert space for the bras by
\begin{equation}
     {\mathbf \Phi}_{\rm exp} \subset L^2([0,\infty ), \rmd r) 
      \subset {\mathbf \Phi}_{\rm exp}^{\prime} \, ,
         \label{rhsexpp}
\end{equation} 
and the one for the kets by
\begin{equation}
     {\mathbf \Phi}_{\rm exp} \subset L^2([0,\infty ), \rmd r) 
      \subset {\mathbf \Phi}_{\rm exp}^{\times} \, .
     \label{rhsexpt}
\end{equation} 

The procedure to construct the space of test functions 
${\mathbf \Phi}_{\rm exp}$ has been explained in~\cite{DIS,LS1,LS2}. The
most important property one has to look at is the ``bad behavior'' of
the Gamow eigenfunctions. Such ``bad behavior'' must be compensated by
the ``nice behavior'' of the elements of ${\mathbf \Phi}_{\rm exp}$ so the
integrals~(\ref{Gketdef})-(\ref{Gbradefb-k}) converge. Since the regular 
solution $\chi (r;q)$ of the Schr\"odinger equation is related to the Gamow 
eigenfunction by
\begin{equation}
      \chi (r;k_n) = \frac{1}{2\rmi} \frac{{\cal J}_-(k_n)}{N_n} u(r;k_n) \, ,
      \quad   n=\pm 1,\pm 2, \ldots \, ,  
\end{equation}
and since by, for example, Eq.~(12.6) in Ref.~\cite{TAYLOR} the
regular solution satisfies
\begin{equation}
      \left| \chi (r;q)\right| \leq C \, 
        \frac{\left|q\right|r}{1+\left|q\right|r} \,  
      \rme ^{|{\rm Im}(q)|r} \, , \quad q\in {\mathbb C} \, ,   
      \label{boundrs}
\end{equation}
the ``bad behavior'' of the Gamow eigenfunctions is given by
\begin{equation}
       \left| u(r;k_n)\right| \leq C \, 
            \frac{|N_n|}{|{\cal J}_-(k_n)|}  \,
            \frac{\left|k_n\right|r}{1+\left|k_n\right|r} \,           
         \rme ^{|{\rm Im}(k_n)|r} \, , \quad 
              n=\pm 1,\pm 2, \ldots  \, .
    \label{estimateofu}
\end{equation}
Because the bound~(\ref{boundrs}) is sharp~\cite{TAYLOR}, so is 
the bound~(\ref{estimateofu}). Thus, the 
Gamow eigenfunctions grow exponentially as $r$ tends
to infinity, and, in order for the integrals~(\ref{Gketdef})-(\ref{Gbradefb-k})
to converge, the wave functions of ${\mathbf \Phi}_{\rm exp}$ must fall off
at infinity sufficiently rapidly.

From Eq.~(\ref{estimateofu}), it is clear that the integrals in
Eqs.~(\ref{Gketdef})-(\ref{Gbradefb-k})
converge already for functions that fall off at infinity faster than any 
exponential~\cite{BOLLINI1,BOLLINI2}. Thus, exponential falloff is the
weakest falloff that we need to require from the wave functions of
$\mathbf{\Phi}_{\rm exp}$, see Refs.~\cite{BOLLINI1,BOLLINI2}. However,
we are going to impose a stronger, Gaussian falloff because it allows 
us to perform certain resonance
expansions, as will be discussed in Sec.~\ref{sec:resexp}.

Using the estimate~(\ref{estimateofu}), and following the procedure
of~\cite{DIS,LS1,LS2} to construct spaces of test functions, one ends
up finding that ${\mathbf \Phi}_{\rm exp}$ is given by
\begin{equation}
      {\mathbf \Phi}_{\rm exp}= 
      \left\{ \varphi \in {\cal D} \, | 
      \ \| \varphi \|_{m,m'}<\infty \, , \ m,m'=0,1,2,\ldots \right\} ,
       \label{phiexp}
\end{equation}
where $\cal D$ is the maximal invariant subspace of the Hamiltonian,
\begin{equation}
       {\cal D} = \bigcap_{m=0}^{\infty} {\cal D}(H^m) \, ,
\end{equation}
and $\| \cdot \|_{m,m'}$ is given by
\begin{equation}
   \hskip-2cm   \| \varphi \|_{m,m'} := \sqrt{\int_{0}^{\infty}\rmd r \, 
    \left| \frac{mr}{1+mr}\, \rme ^{mr^2/2} (1+H)^{m'}
             \varphi (r) \right|^2 \, } 
                 \, , \quad m,m'=0,1,2, \ldots \, .
      \label{normsLS}
\end{equation}
Hence, ${\mathbf \Phi}_{\rm exp}$ is just the space of square integrable 
functions which belong to the
maximal invariant subspace of $H$ and for which the quantities~(\ref{normsLS})
are finite. In particular, because $\varphi (r)$ satisfies the 
estimates~(\ref{normsLS}), $\varphi (r)$ falls off at infinity faster 
than $\rme ^{-r^2}$, that is, its tails fall off faster than Gaussians. 

Note that we have arrived at the same space of test functions as the one
for the analytically continued Lippmann-Schwinger bras and kets~\cite{LS2}, 
since also in that case we have to tame real exponentials.

Once we have constructed the space $\mathbf \Phi _{\rm exp}$, we can 
construct its dual $\mathbf \Phi _{\rm exp}^{\prime}$ and antidual
$\mathbf \Phi _{\rm exp}^{\times}$ spaces as the 
spaces of, respectively, linear and antilinear continuous 
functionals over $\mathbf \Phi _{\rm exp}$, and therewith 
the rigged Hilbert spaces~(\ref{rhsexpp}) and (\ref{rhsexpt}). The 
Gamow bras and kets are, respectively, linear and antilinear continuous 
functionals over ${\mathbf \Phi}_{\rm exp}$. As well, they are (generalized)
eigenvectors of the Hamiltonian.

The following proposition, whose proof follows exactly the same steps as the 
proof of Proposition~2 in~\cite{LS2}, encapsulates the results of this section:

\vskip0.5cm

\newtheorem*{Prop1}{Proposition~1}
\begin{Prop1} \label{Prop1} The triplets of spaces~(\ref{rhsexpp}) and 
(\ref{rhsexpt}) are rigged Hilbert spaces, and they satisfy all
the requirements to accommodate the Gamow bras and kets. More specifically,
\begin{itemize}

\item[({\it i})] The $\| \cdot \|_{m,m'}$ are norms, and they
define a countably normed topology, i.e., a meaning of sequence convergence.

\item[({\it ii})] The space ${\mathbf \Phi}_{\rm exp}$ is dense in
$L^2([0,\infty ),\rmd r)$.

\item[({\it iii})] The space ${\mathbf \Phi}_{\rm exp}$ is invariant under
the action of the Hamiltonian, and $H$ is 
${\mathbf \Phi _{\rm exp}}$-continuous.

\item[({\it iv})] The kets $|z_n\rangle$ are continuous, 
{\it antilinear} functionals over ${\mathbf \Phi}_{\rm exp}$, i.e., 
$|z_n\rangle \in {\mathbf \Phi}_{\rm exp}^{\times}$.

\item[({\it v})] The kets $|z_n\rangle$ are generalized ``right'' 
eigenvectors of $H$ with eigenvalue $z_n$:
\begin{equation}
      H|z_n\rangle= z_n \, |z_n\rangle \, , \quad n=\pm 1,\pm 2, \ldots  \, ;
       \label{keigeeqa}
\end{equation}
that is,
\begin{equation}
      \langle \varphi |H|z_n\rangle =
         z_n \langle \varphi |H|z_n\rangle \, , \quad 
       \varphi  \in {\mathbf \Phi}_{\rm exp}  \, .
      \label{keigeeqbis} 
\end{equation}

\item[({\it vi})] The bras $\langle z_n|$ are continuous, 
{\it linear} functionals over ${\mathbf \Phi}_{\rm exp}$, i.e., 
$\langle z_n| \in {\mathbf \Phi}_{\rm exp}^{\prime}$.

\item[({\it vii})]The bras $\langle z_n|$ are generalized ``left'' 
eigenvectors of $H$ with eigenvalue $z_n$:
\begin{equation}
      \langle z_n|H= z_n \langle z_n| \, , \quad n=\pm 1,\pm 2, \ldots  \, ;
             \label{kpssleftkeofHa}
\end{equation}
that is,
\begin{equation}
       \langle z_n|H|\varphi \rangle 
       = z_n \langle z_n| \varphi  \rangle  \, , \quad 
       \varphi  \in {\mathbf \Phi}_{\rm exp}  \, .
        \label{kpssleftkeofHb}   
\end{equation}
\end{itemize}
\end{Prop1}

\vskip0.5cm

Proposition~1 makes it clear, in particular, that there is a 1:1 
correspondence between Gamow bras and kets. 

Note that in terms of the wave number, the eigenequations~(\ref{keigeeqa}) 
and (\ref{kpssleftkeofHa}) become
\begin{equation}
      H|k_n\rangle= \frac{\hbar ^2}{2m}k_n^2 \, |k_n\rangle 
        \, , \quad n=\pm 1,\pm 2, \ldots \, ,
       \label{keigeeqaq}
\end{equation}
\begin{equation}
      \langle k_n|H=\frac{\hbar ^2}{2m} k_n^2  \langle k_n|
        \, , \quad n=\pm 1,\pm 2, \ldots \, .
             \label{kpssleftkeofHaq}
\end{equation}
Note also that the bra eigenequation~(\ref{kpssleftkeofHa}) is \emph{not} 
given by
\begin{equation}
       \langle z_n|H = z_n^*\langle z_n| \, ,
          \label{naiveignd} 
\end{equation}
as one may naively obtain by Hermitian conjugation
of the ket eigenequation~(\ref{keigeeqa}). The reason
lies in that one has to use complex Hermitian conjugation to obtain
the ``left'' from the ``right'' Gamow eigenfunction, see 
Eq.~(\ref{legazn}). 

The normalization condition satisfied by the Gamow states
is the following:
\begin{equation}
       \langle z_n|z_{n'} \rangle = \delta _{n,n'} \, , 
           \quad n,n'=\pm 1, \pm 2 , \ldots  \, .
              \label{orthGS} 
\end{equation}
When $n=n'$, Eq.~(\ref{orthGS}) follows from Zeldovich's 
regularization~(\ref{Zregula}). When $n\neq n'$, Eq.~(\ref{orthGS}) can be
proved in the same way as one proves the orthogonality of bound states:
\begin{equation}
       \langle z_n|H|z_{n'} \rangle = z_n \langle z_n|z_{n'} \rangle =
       z_{n'} \langle z_n|z_{n'} \rangle \, ,
            \label{indsfjannp}
\end{equation}
where we have made use of the fact that $\langle z_n|$ and $|z_{n'} \rangle$
are eigenvectors of $H$ with eigenvalue $z_n$ and $z_{n'}$, respectively. From
the second equality in~(\ref{indsfjannp}), we obtain
\begin{equation}
       (z_n - z_{n'}) \langle z_n|z_{n'} \rangle =0 \, ,
            \label{indsfjannpznd}
\end{equation}
which yields the desired result, since $z_n \neq z_{n'}$ when $n\neq n'$.
It should be noted however that, similar to the normalization of
scattering states, the normalization condition~(\ref{orthGS})
has only formal meaning and does not imply the use of a Hilbert-space 
scalar product. For example, the ``scalar product'' built on 
Eqs.~(\ref{orthGS}) and~(\ref{Zregula}) would not satisfy
$(f,f)\geq 0$.

\section{The energy representations of rigged Hilbert spaces
and of the Gamow bras and kets}
\label{sec:Gvecenwnrepr}

We turn now to obtain and characterize the energy representations of 
the rigged Hilbert spaces~(\ref{rhsexpp}) and (\ref{rhsexpt}) and of 
the Gamow bras and kets. It is here where we will need to introduce 
the labels $\pm$ in the notation for the wave functions and for the
Gamow bras and kets. 

\subsection{The energy representations of the rigged Hilbert space}

The ``in'' and the ``out'' energy representations of 
${\mathbf \Phi}_{\rm exp}$ are readily obtained by means of the unitary 
operators $U_{\pm}$ of~\cite{LS1}:
\begin{equation}
     {U}_{\pm}{\mathbf \Phi}_{\rm exp} \equiv  
          \widehat{\mathbf \Phi}_{\pm {\rm exp}} \, ,
\end{equation}
which in turn yield the energy representations of the rigged Hilbert
spaces~(\ref{rhsexpp}) and~(\ref{rhsexpt}):
\begin{equation}
     \widehat{\mathbf \Phi}_{\pm {\rm exp}} \subset L^2([0,\infty ),\rmd E)
      \subset \widehat{\mathbf \Phi}_{\pm {\rm exp}}^{\prime}   \, ,
\end{equation}
\begin{equation}
     \widehat{\mathbf \Phi}_{\pm {\rm exp}} \subset L^2([0,\infty ),\rmd E)
      \subset \widehat{\mathbf \Phi}_{\pm {\rm exp}}^{\times}   \, .
\end{equation}
The elements of $\widehat{\mathbf \Phi}_{\pm {\rm exp}}$ will be denoted
by $\widehat{\varphi}^{\pm}(z)= U_{\pm}\varphi (z)$. 

In~\cite{LS2}, we characterized the analytic and growth properties of the wave 
functions in the wave number representations, $\widehat{\varphi}^{\pm}(q)$,
which are related to the wave functions in the energy representations as
\begin{equation}
     \widehat{\varphi}^{\pm}(z)  = \sqrt{\frac{2m}{\hbar ^2} \frac{1}{2q} \,} 
        \,  \widehat{\varphi}^{\pm}(q) \, .
\end{equation}
Thus, the results of~\cite{LS2} also characterize the analytic and growth 
properties of $\widehat{\varphi}^{\pm}(z)$, and we will refer to~\cite{LS2}
whenever we need to make use of any such properties.

As mentioned above, from now on we will add a label to the action
of the Gamow states,
\begin{equation}
        \langle \varphi ^{\pm}|z_n^{\pm}\rangle \, , \quad 
        \langle ^{\pm}z_n|\varphi ^{\pm}\rangle \, ,  \qquad 
        n=\pm 1, \pm 2, \ldots  \, .
\end{equation}
When we use the label $+$, it will mean that the energy representation is
obtained through the operator $U_+$, and when we use the label $-$, it 
will mean that the energy representation is obtained through the 
operator $U_-$.

\subsection{The energy representations of the Gamow bras and kets}

In order to obtain the energy representations of the Gamow bras and kets, 
we first need to define the linear complex delta functional at $z$:
\begin{equation}
          \hskip-1cm
     \begin{array}{rcl}
           \langle \widehat{\delta}_z| : \widehat{\mathbf \Phi}_{\rm exp} & 
                \longmapsto & {\mathbb C}  \nonumber \\ 
            \widehat{\varphi} & \longmapsto & 
                     \langle \widehat{\delta}_z| \widehat{\varphi} \rangle :=
                       \widehat{\varphi}(z)  \, ,
                     \label{lincdfcun}
     \end{array}
\end{equation}
where $\widehat{\mathbf \Phi}_{{\rm exp}}$ may be either 
$\widehat{\mathbf \Phi}_{+{\rm exp}}$ or 
$\widehat{\mathbf \Phi}_{-{\rm exp}}$, and $\widehat{\varphi}$ may be either
$\widehat{\varphi}^+$ or $\widehat{\varphi}^-$. Thus, the linear complex delta
functional at $z$ associates with each test function, the value of the test 
function at $z$. One can write~(\ref{lincdfcun}) as an integral operator
as
\begin{equation}
     \langle \widehat{\delta}_z| \widehat{\varphi} \rangle =
      \int_0^{\infty}\rmd E \, \delta (E-z) \widehat{\varphi}(E) =
            \widehat{\varphi}(z)  \, .
\end{equation}
In this way, one can interpret the complex delta function $\delta (E-z)$ 
as the analytic continuation of the Dirac delta function $\delta (E-E')$.

The antilinear complex delta functional 
$| \widehat{\delta}_z\rangle$ at the complex number $z$ can be defined 
in a similar way:
\begin{equation}
          \hskip-1cm
     \begin{array}{rcl}
           |\widehat{\delta}_z \rangle : \widehat{\mathbf \Phi}_{\rm exp} & 
                \longmapsto & {\mathbb C}  \nonumber \\ 
            \widehat{\varphi} & \longmapsto & 
                     \langle \widehat{\varphi}| \widehat{\delta}_z \rangle :=
                   [\widehat{\varphi}(z^*)]^*  \, .
                     \label{antcdfcun}
     \end{array}
\end{equation}
We also need to define the linear and antilinear residue functionals at $z$:
\begin{equation}
          \hskip-1cm
     \begin{array}{rcl}
           \langle \widehat{\rm res}_z| : \widehat{\mathbf \Phi}_{\rm exp} & 
                \longmapsto & {\mathbb C}  \nonumber \\ 
            \widehat{\varphi} & \longmapsto & 
                     \langle \widehat{\rm res}_z| \widehat{\varphi} \rangle :=
                       {\rm res}[\widehat{\varphi}(z)]  \, ,
                     \label{lincresfcun}
     \end{array}
\end{equation}
\begin{equation}
          \hskip-1cm
     \begin{array}{rcl}
           | \widehat{\rm res}_z \rangle : \widehat{\mathbf \Phi}_{\rm exp} & 
                \longmapsto & {\mathbb C}  \nonumber \\ 
            \widehat{\varphi} & \longmapsto & 
                     \langle \widehat{\varphi}|\widehat{\rm res}_z \rangle :=
                   {\rm res}[\widehat{\varphi}(z^*)]^*  \, ,
                     \label{antcresfcun}
     \end{array}
\end{equation}
where ${\rm res}[\widehat{\varphi}(z)]$ stands for the residue of 
$\widehat{\varphi}$ at $z$. Both the complex delta functionals and the residue 
functionals at $z$ are well defined when the test functions can be analytically
continued into $z$, as is our case~\cite{LS2}.

We will need also the following normalization factor:
\begin{equation}
      {\cal N}_n^2=\rmi \, {\rm res}[S(z)]_{z=z_n} = 
        \rmi \frac{\hbar ^2}{2m}2k_n \, {\rm res}[S(q)]_{q=k_n} = 
        \frac{\hbar ^2}{2m}2k_n N_n^2 \, ,
\end{equation}
where $N_n$ was used in Sec.~\ref{sec:gamowvectors} to normalize the Gamow 
eigenfunctions. 

If we denote the energy representations of the Gamow bras and kets as
\begin{equation}
      \langle ^{\pm}\widehat{z}_n| \equiv 
         \langle ^{\pm} z_n|U_{\pm} \, ,
\end{equation}
\begin{equation}
      |\widehat{z}_n^{\pm}\rangle \equiv 
       U_{\pm}|z_n^{\pm}\rangle \, ,
\end{equation}
then the following proposition, whose proof can be found 
in Appendix~\ref{sec:proofsofprop}, holds:

\vskip0.5cm

\newtheorem*{Prop2}{Proposition~2}
\begin{Prop2} \label{Prop2} For a resonance (or anti-resonance) of energy 
$z_n$, the ``minus'' (or ``out'') energy representation of the Gamow 
bras and kets is given by
\begin{equation}
     \langle ^-\widehat{z}_n| = 
        -  \frac{\sqrt{2\pi \,}}{{\cal N}_n} \,
     \langle \widehat{\rm res}_{z_n}|  \, , \quad  n=\pm 1, \pm 2, \ldots \, ,
      \label{-ErepGb}
\end{equation}
\begin{equation}
     |\widehat{z}_n^- \rangle = \rmi \sqrt{2\pi \, } {\cal N}_n \,
      |\widehat{\delta}_{z_n} \rangle  \, , \quad  n=\pm 1, \pm 2, \ldots \, .
      \label{-ErepGk}
\end{equation}
Their ``plus'' energy representation is given by 
\begin{equation}
       \langle ^+ \widehat{z}_n| = 
        \rmi \sqrt{2\pi \, } {\cal N}_n \,
        \langle \widehat{\delta}_{z_n}|  \, , 
            \quad  n=\pm 1, \pm 2, \ldots \, ,
         \label{+ErepGb}
\end{equation}
\begin{equation}
     |\widehat{z}_n^+ \rangle = 
        - \frac{\sqrt{2\pi \, }}{{\cal N}_n} \,
      |\widehat{\rm res}_{z_n} \rangle  \, , \quad  n=\pm 1, \pm 2, \ldots \, .
      \label{+ErepGk}
\end{equation}
\end{Prop2}

\vskip0.5cm

Proposition~2 shows, in particular, that the ``plus'' energy representation of
a Gamow bra or ket is different from its ``minus'' energy representation,
thereby showing that the labels $\pm$ matter.

The complex delta functional and the residue functional can be written
in more familiar terms as follows. By using the resolutions of 
the identity
\begin{equation}
       I = \int_0^{\infty}\rmd E \, |E^{\pm} \rangle \langle ^{\pm} E| \, ,  
     \label{residenpm}
\end{equation}
we can formally write the actions of 
$\langle ^{\pm}\widehat{z}_n|$ as integral operators and obtain
\begin{eqnarray}
     \langle ^{\pm}\widehat{z}_n| \widehat{\varphi}^{\pm} \rangle &=&     
      \langle ^{\pm}z_n|{\varphi}^{\pm} \rangle  \nonumber \\
     & =& \int_0^{\infty}\rmd E \, 
      \langle ^{\pm}z_n|E^{\pm}\rangle \langle ^{\pm}E|{\varphi}^{\pm} \rangle
     \nonumber \\
      & =& \int_0^{\infty}\rmd E \, 
      \langle ^{\pm}z_n|E^{\pm}\rangle \, \widehat{\varphi}^{\pm} (E)  \, .
        \label{braqpmintope}
\end{eqnarray}
Comparison of (\ref{braqpmintope}) with (\ref{-ErepGb}) and (\ref{+ErepGb})
shows that $\langle ^-z_n|E^-\rangle$ is proportional to the residue
distribution,
\begin{equation}
     \langle ^-z_n|E^-\rangle = 
            - \frac{\sqrt{2\pi \, }}{{\cal N}_n} \,
       {\rm res}[ \, \cdot \, ]_{z_n}    \, , \qquad
       E\geq 0 \, , \quad  n=\pm 1, \pm2 , \ldots  \, ,
\end{equation}
and that $\langle ^+z_n|E^+\rangle$ is proportional to the complex delta
function,
\begin{equation}
      \langle ^+z_n|E^+\rangle =
       \rmi \sqrt{2\pi \, } {\cal N}_n \,  \delta (E-z_n) \, , \qquad
       E\geq 0 \, , \quad  n=\pm 1, \pm2 , \ldots \, .
          \label{brapzcd}
\end{equation}

Similarly, by using~(\ref{residenpm}) we can formally write the actions of 
$|\widehat{z}_n^{\pm}\rangle$ as integral operators:
\begin{eqnarray}
     \langle  \widehat{\varphi}^{\pm}|\widehat{z}_n^{\pm} \rangle &=&     
      \langle {\varphi}^{\pm}|z_n^{\pm} \rangle  \nonumber \\
     & =& \int_0^{\infty}\rmd E \, 
     \langle {\varphi}^{\pm}|E^{\pm}\rangle \langle ^{\pm}E|z_n^{\pm} \rangle
     \nonumber \\
      & =& \int_0^{\infty}\rmd E \, 
      [\varphi ^{\pm}(E)]^* \langle ^{\pm}E|z_n^{\pm} \rangle \, .
        \label{acwpmq}
\end{eqnarray}   
By comparing (\ref{acwpmq}) with (\ref{-ErepGk}) and (\ref{+ErepGk}), we
deduce that $\langle ^-E|z_n^-\rangle$ is proportional to the complex delta
function
\begin{equation}
        \langle ^-E|z_n^-\rangle =
       \rmi \sqrt{2\pi \, } {\cal N}_n \,  \delta (E-z_n) \, , \qquad
       E\geq 0 \, , \quad  n=\pm 1, \pm2 , \ldots \, ,
          \label{ketpzcd}
\end{equation}
and that $\langle ^+E|z_n^+\rangle$ is proportional to the residue
distribution,
\begin{equation}
        \langle ^+E|z_n^+\rangle = 
       - \frac{\sqrt{2\pi \,}}{{\cal N}_n} \,
       {\rm res}[ \, \cdot \, ]_{z_n}    \, , \qquad
       E\geq 0 \, , \quad  n=\pm 1, \pm2 , \ldots  \, .
\end{equation}

It is important to realize that with a given test function, the complex delta 
function and the residue distribution at $z_n$ associate, respectively, the 
value and the residue of the analytic continuation of the test function at 
$z_n$. This is why when those distributions act on 
$[\widehat{\varphi}^{\pm}(E)]^*$ as in Eq.~(\ref{acwpmq}), the final result 
is respectively $[\widehat{\varphi}^{-}(z_n^*)]^*$ and
${\rm res} \, [ \widehat{\varphi}^{+}(z_n^*)]^*$, rather
than $[\widehat{\varphi}^{-}(z_n)]^*$ and
${\rm res} \, [ \widehat{\varphi}^{+}(z_n)]^*$, since 
the analytic continuation of $[\widehat{\varphi}^{\pm}(E)]^*$ is
$[\widehat{\varphi}^{\pm}(z^*)]^*$ rather than
$[\widehat{\varphi}^{\pm}(z)]^*$. 



\section{The $(-\infty , \infty )$-``energy'' representation}
\label{sec:theminpinerep}

The spectrum of our Hamiltonian is $[0,\infty )$. Hence, in 
Eqs.~(\ref{brapzcd}) and (\ref{ketpzcd}) the energy $E$ runs over the positive
real line. In this section, we are going to let $E$ run over
the full real line. In doing so, we can see what would happen if the
spectrum of the Hamiltonian wasn't bounded from below. 


It is important to keep in mind that in this section, we will need to treat 
resonances and anti-resonances separately. Also, strictly speaking, whenever 
we say that $E$ runs over over the full real line $(-\infty ,\infty )$, it 
will actually mean that in the case of
resonances (anti-resonances), $E$ runs infinitesimally below (above)
the real axis of the second sheet of the Riemann surface.

In order to construct the $(-\infty , \infty)$-``energy'' representation, we
will construct the transform $\mathbb A$ that lets the energy vary over
the full real line. The transform $\mathbb A$ is modeled after the
``$\theta$ transform'' of~\cite{BG}, and it allows us to connect the
physical spectrum, which in our case coincides with $[0,\infty)$, with the 
support of the Breit-Wigner amplitude, which coincides with 
$(-\infty , \infty)$. Basically, $\mathbb A$ takes a test
function $\widehat{\varphi}^{\pm}(E)$, $E\geq 0$, of
$\widehat{\mathbf \Phi}_{\pm{\rm exp}}$ into its analytic continuation
over the full real line, $\widehat{\varphi}^{\pm}(E)$, 
$E\in (-\infty ,\infty)$. In order to distinguish when the energy runs 
over the physical spectrum from when it runs over the full real line, we 
will denote $\widehat{\varphi}^{\pm}(E)$, $E\in (-\infty ,\infty)$, by 
$\widetilde{\varphi}^{\pm}(E)$ and thus will write
\begin{equation}
       {\mathbb A}\widehat{\varphi}^{\pm}\equiv \widetilde{\varphi}^{\pm} \, , 
\end{equation}
and
\begin{equation}
     {\mathbb A} \widehat{\mathbf \Phi}_{\pm{\rm exp}}\equiv
     \widetilde{\mathbf \Phi}_{\pm{\rm exp}} =
    \{ \widetilde{\varphi}^{\pm}(E) \ | \quad E\in (-\infty , \infty) \} \, .
\end{equation}

The following diagram shows how ${\mathbb A}$ links the energy representation 
with the $(-\infty ,\infty )$-``energy'' representation:
\begin{equation}
       \hskip-1.7cm   
    \begin{array}{rcccccccc}
       \widehat{\varphi}^{\pm} \, , \ &  
      \widehat{\mathbf \Phi}_{{\pm}{\rm exp}} & 
       \subset & 
      L^2([0,\infty),\rmd E) & \subset & 
      \widehat{\mathbf \Phi}_{{\pm}{\rm exp}}^{\times} & 
      \  & \mbox{energy representation}  \\ [1ex]
       \ & 
      \downarrow {\mathbb A} & 
       \ & \ & \ & \downarrow {\mathbb A}  & 
      \ & \  \\ [1ex]
       \widetilde{\varphi}^{\pm} \, , \ & 
       \widetilde{\mathbf \Phi}_{{\pm}{\rm exp}} & 
       \subset & 
       L^2({\mathbb R}, \rmd _{\alpha}E) & \subset & 
      \widetilde{\mathbf \Phi}_{{\pm}{\rm exp}}^{\times} & 
      \ & 
      (-\infty ,\infty )\mbox{-``energy'' repr.}  \\
     \end{array}
      \label{thetafuncis}
\end{equation}
where $L^2({\mathbb R}, \rmd _{\alpha}E)$ is the following space: 
\begin{equation}
     \hskip-1.5cm
     L^2({\mathbb R}, \rmd _{\alpha}E) = \{ \widetilde{f} \ | \quad  
             \widehat{f}\in L^2([0,\infty ),\rmd E) \, , \
   \lim_{\alpha \to 0}\int_{-\infty}^{\infty}\rmd E\, |\rme ^{-\rmi E \alpha}
             \widetilde{f}(E)|^2 < \infty \} \, .
     \label{l2alpha}
\end{equation} 
In Eq.~(\ref{l2alpha}), the integral is assumed to be calculated in the
second sheet, infinitesimally below (or above, in the case of anti-resonances)
the real axis. The convergence factor 
$\rme ^{-\rmi E\alpha}$ (which becomes $\rme ^{\rmi E\alpha}$ in the case
of anti-resonances) is needed because the analytic 
continuation of $\widehat{\varphi}^{\pm}(E)$ into the negative energies
blows up exponentially~\cite{LS2}. Actually, if it wasn't needed, 
the spectrum would be the full real line. Nevertheless, the space
$L^2({\mathbb R}, \rmd _{\alpha}E)$ is not crucial to our discussion. 

It is important to understand that although we have denoted the functions
$\widehat{\varphi}^{\pm}$ and 
$\widetilde{\varphi}^{\pm}={\mathbb A}\widehat{\varphi}^{\pm}$ by a different 
symbol, they are indeed the {\it same} function. More precisely, they are 
different ``pieces'' of the same function. In particular, the value of their 
analytic continuation at a complex number $z$ is the same,
\begin{equation}
       \widetilde{\varphi}^{\pm}(z) = \widehat{\varphi}^{\pm}(z) \, .
\end{equation}
Obviously, the analytic continuation of their complex conjugates enjoys
an analogous property,
\begin{equation}
       [\widetilde{\varphi}^{\pm} (z^*)]^* = 
      [\widehat{\varphi}^{\pm} (z^*)]^*  \, .
       \label{samefunction} 
\end{equation}
We use different symbols for different ``pieces'' of the same
function because the proof of the connection between the Breit-Wigner 
amplitude and the complex delta function becomes more transparent. For 
resonances, such connection is given by
\begin{equation}
      {\mathbb A}|\widehat{z}_n^-\rangle = |\frac{1}{E-z_n}{^-} \rangle \, ,
        \quad n=1,2, \ldots \, ,
      \label{bwcomfrel}
\end{equation}
where the ket $|\frac{1}{E-z_n}{^-} \rangle$ is associated with the 
Breit-Wigner amplitude as follows:
\begin{equation}
      \hskip-2.6cm 
  \begin{array}{rcl}
      |\frac{1}{E-z_n}{^-} \rangle :\widetilde{\mathbf \Phi}_{-{\rm exp}} 
      & \mapsto & \mathbb C \nonumber \\
         \widetilde{\varphi}^- & \mapsto & 
       \langle \widetilde{\varphi}^-|\frac{1}{E-z_n}{^-} \rangle :=
    \lim_{\alpha \to 0}\int_{-\infty}^{\infty}\rmd E\, \rme ^{-\rmi E \alpha} 
         \left( -\frac{{\cal N}_n}{\sqrt{2\pi}} \frac{1}{E-z_n} \right)
        [\widetilde{\varphi}^-(E)]^* \, .
        \label{BWfunction}
   \end{array} 
\end{equation}
We will call this ket the Breit-Wigner ket. The integral in
Eq.~(\ref{BWfunction}) is supposed to be calculated in the lower half plane
of the second sheet, infinitesimally below the real axis. By the properties of 
$\widehat{\varphi}^-(z)$ in the lower half plane of the second 
sheet~\cite{LS2}, the Breit-Wigner ket is a well defined antilinear 
functional. The proof of~(\ref{bwcomfrel}) is provided 
in Appendix~\ref{sec:proofsofprop}.

The combination of~(\ref{bwcomfrel}) with the results of 
Sec.~\ref{sec:Gvecenwnrepr} shows that the Gamow eigenfunction 
$u(r;z_n)$, the complex
delta function (multiplied by a normalization factor) and
the Breit-Wigner amplitude (multiplied by a normalization factor) 
are the same distribution in different representations:
\begin{equation}
         \hskip-2.7cm
    \begin{array}{ccccc}
       u(r;z_n)& \leftrightarrow  & 
      \rmi \sqrt{2\pi \, } {\cal N}_n \delta (E-z_n) \, , \ E\in [0,\infty ) & 
      \leftrightarrow  & -\frac{{\cal N}_n}{\sqrt{2\pi}}\,
       \frac{1}{E-z_n}\, , \ E\in (-\infty ,\infty)  \\ [2ex]
       \mbox{posit. repr.} &\ &  \mbox{energy repr.} & \ & 
      (-\infty,\infty)\mbox{-``energy'' repr.}
    \end{array}
\end{equation}
Physically, these links mean that the Gamow states yield a decay amplitude 
given by the complex delta function, and that such decay amplitude can be 
approximated 
by the Breit-Wigner amplitude when we can ignore the lower bound of the
energy, i.e., when the resonance is so far from the threshold that we
can safely assume that the energy runs over the full real 
line. However, because there is actually a lower bound for the energy, the 
decay amplitude is never exactly given by the Breit-Wigner 
amplitude. Mathematically,
the reason lies in that ${\mathbb A}$ is not unitary, which makes the
energy representation be not equivalent to the 
$(-\infty,\infty)$-``energy'' representation.

One can also relate the ``plus'' Gamow bra with a Breit-Wigner bra: 
\begin{equation}
       \langle ^+\widehat{z}_n|{\mathbb A} = \langle ^+\frac{1}{E-z_n}| \, ,
            \quad n=1,2, \ldots \, ,
     \label{BWgpbr}
\end{equation}
where the Breit-Wigner bra is defined as
\begin{equation}
      \hskip-2.5cm 
  \begin{array}{rcl}
      \langle ^+\frac{1}{E-z_n}| :\widetilde{\mathbf \Phi}_{+{\rm exp}} 
      & \mapsto & \mathbb C \nonumber \\
         \widetilde{\varphi}^+ & \mapsto & 
       \langle ^+\frac{1}{E-z_n}|\widetilde{\varphi}^+\rangle :=
    \lim_{\alpha \to 0}\int_{-\infty}^{\infty}\rmd E\, \rme ^{-\rmi E \alpha} 
         \left( -\frac{{\cal N}_n}{\sqrt{2\pi}} \frac{1}{E-z_n} \right)
        \widetilde{\varphi}^+(E) \, .
        \label{BWfunctionbra}
   \end{array} 
\end{equation}
The proof of~(\ref{BWgpbr}) is almost identical to the proof 
of~(\ref{bwcomfrel}). 

For the anti-resonance energies, we obtain similar results to~(\ref{bwcomfrel})
and (\ref{BWgpbr}). The Gamow ket of an anti-resonance is related to a 
Breit-Wigner ket as
\begin{equation}
      {\mathbb A}|\widehat{z}_{n}^-\rangle = 
             |\frac{1}{E-z_{n}}{^-} \rangle \, ,
        \quad n=-1,-2, \ldots \, ,
      \label{bwcomfrelanti}
\end{equation}
where now the ket $|\frac{1}{E-z_{n}}{^-} \rangle$ is associated with the 
Breit-Wigner amplitude as follows:
\begin{equation}
      \hskip-2.5cm 
  \begin{array}{rcl}
      |\frac{1}{E-z_{n}}{^-} \rangle :\widetilde{\mathbf \Phi}_{-{\rm exp}} 
      & \mapsto & \mathbb C \nonumber \\
         \widetilde{\varphi}^- & \mapsto & 
       \langle \widetilde{\varphi}^-|\frac{1}{E-z_{n}}{^-} \rangle :=
    \lim_{\alpha \to 0}\int_{-\infty}^{\infty}\rmd E\, \rme ^{\rmi E \alpha} 
         \left( \frac{{\cal N}_n}{\sqrt{2\pi}} \frac{1}{E-z_{n}} \right)
        [\widetilde{\varphi}^-(E)]^* \, .
        \label{BWfunctionanti}
   \end{array} 
\end{equation}
Similarly, the Gamow bra is associated with a Breit-Wigner bra as
\begin{equation}
       \langle ^+\widehat{z}_n|{\mathbb A} = \langle ^+\frac{1}{E-z_n}| \, ,
            \quad n=-1,-2, \ldots \, ,
     \label{BWgpbranti}
\end{equation}
where the Breit-Wigner bra is now defined as
\begin{equation}
      \hskip-2cm 
  \begin{array}{rcl}
      \langle ^+\frac{1}{E-z_n}| :\widetilde{\mathbf \Phi}_{+{\rm exp}} 
      & \mapsto & \mathbb C \nonumber \\
         \widetilde{\varphi}^+ & \mapsto & 
       \langle ^+\frac{1}{E-z_n}|\widetilde{\varphi}^+\rangle :=
    \lim_{\alpha \to 0}\int_{-\infty}^{\infty}\rmd E\, \rme ^{\rmi E \alpha} 
         \left( \frac{{\cal N}_n}{\sqrt{2\pi}} \frac{1}{E-z_n} \right)
        \widetilde{\varphi}^+(E) \, .
        \label{BWfunctionbraanti}
   \end{array} 
\end{equation}
The proofs of~(\ref{bwcomfrelanti}) and (\ref{BWgpbranti}) are
very similar to those of~(\ref{bwcomfrel}) and (\ref{BWgpbr}). In 
Eqs.~(\ref{BWfunctionanti}) and (\ref{BWfunctionbraanti}), the integration 
is supposed to be done infinitesimally {\it above} the real axis of the 
second sheet, contrary to Eqs.~(\ref{BWfunction}) and (\ref{BWfunctionbra}), 
where the integration is supposed to be done infinitesimally {\it below} 
the real axis of the second sheet. Also, in Eqs.~(\ref{BWfunctionanti})
and (\ref{BWfunctionbraanti}) the regulator is $\rme ^{\rmi E \alpha}$, 
$\alpha >0$, whereas in Eqs.~(\ref{BWfunction}) and (\ref{BWfunctionbra}) the 
regulator is $\rme ^{-\rmi E \alpha}$, $\alpha >0$. The reason why 
anti-resonances need the opposite sign in their regulator will become 
apparent in Sec.~\ref{sec:resexp}.

Note that unlike $|z_n^-\rangle$ and $\langle ^+z_n|$,
the ``plus'' Gamow ket $|z_n^+\rangle$
and the ``minus'' Gamow bra $\langle ^-z_n|$ are not related to a
Breit-Wigner amplitude in an obvious way.

The relation between the various representations we have constructed can be
conveniently summarized in diagrams. For resonances we have
\begin{equation}
     \hskip-2.7cm
      \begin{array}{ccccccl}
      H;\, \varphi ^{-}(r) \ & \mathbf \Phi _{\rm exp} & \subset & 
      L^2([0,\infty),\rmd r) &
      \subset & \mathbf \Phi _{\rm exp}^{\times} & 
      \langle r|z_n^-\rangle \equiv u(r;z_n) \nonumber \\  [2ex]
      \ & \downarrow U_- &  &\downarrow U_-  & \
       & \downarrow U_- & \   \nonumber \\ [2ex]  
      \widehat{H}; \, \widehat{\varphi}^-(E) \ & 
      \widehat{\mathbf \Phi}_{-{\rm exp}} & 
       \subset & 
      L^2([0,\infty),\rmd E) & \subset & 
      \widehat{\mathbf \Phi}_{-{\rm exp}}^{\times} & \
       \langle ^-E|z_n^-\rangle \equiv 
               \rmi \sqrt{2\pi}{\cal N}_n\delta (E-z_n) 
       \nonumber  \\ [2ex]  
       & \downarrow {\mathbb A} & \  & \  &
       & \downarrow {\mathbb A} &  \  \nonumber \\ [2ex] 
      \widetilde{H}; \, \widetilde{\varphi}^-(E) \ & 
      \widetilde{\mathbf \Phi}_{-{\rm exp}} & \subset & 
       L^2({\mathbb R},\rmd _{\alpha}E) & \subset & 
        \widetilde{\mathbf \Phi}_{-{\rm exp}}^{\times} 
      &  \langle ^-E|z_n^- \rangle \equiv
      -\frac{{\cal N}_n}{\sqrt{2\pi}} \, \frac{1}{E-z_n} 
       \nonumber   \\
       \end{array}
      \label{rsonsancdiagra} \nonumber
\end{equation}

\vskip.5cm

\noindent and

\vskip.5cm

\begin{equation}
     \hskip-2.7cm
      \begin{array}{ccccccl}
      H; \, \varphi ^{+}(r) \ & \mathbf \Phi _{\rm exp} & \subset & 
      L^2([0,\infty),\rmd r) &
      \subset & \mathbf \Phi _{\rm exp}^{\times} &  
      \langle ^+z_n|r\rangle \equiv u(r;z_n)  \\  [2ex]
      \ &  \downarrow U_+ &  &\downarrow U_+  & \
       & \downarrow U_+ &  \   \\ [2ex]  
      \widehat{H}; \, \widehat{\varphi}^+(E) &  
      \widehat{\mathbf \Phi}_{+{\rm exp}} & 
       \subset & 
      L^2([0,\infty),\rmd E) & \subset & 
      \widehat{\mathbf \Phi}_{+{\rm exp}}^{\times} & 
      \langle ^+z_n|E^+\rangle \equiv 
               \rmi \sqrt{2\pi}{\cal N}_n\delta (E-z_n)  \\ [2ex]  
       & \  \downarrow {\mathbb A} & \ & \  &
       & \downarrow {\mathbb A} &  \   \\ [2ex] 
      \widetilde{H}; \, \widetilde{\varphi}^+(E) &  
      \widetilde{\mathbf \Phi}_{+{\rm exp}} & \subset & 
       L^2({\mathbb R},\rmd _{\alpha}E) & \subset & 
        \widetilde{\mathbf \Phi}_{+{\rm exp}}^{\times} 
      & \langle ^+z_n|E^+ \rangle \equiv
      -\frac{{\cal N}_n}{\sqrt{2\pi}} \, \frac{1}{E-z_n}  \\
       \end{array}
         \label{seoncdig}  \nonumber
\end{equation}
where $\widehat{H}$ denotes the operator multiplication by $E$, $E\geq 0$, and
$\widetilde{H}$ denotes the operator multiplication by $E$, 
$-\infty < E < \infty$. The top, middle and bottom rows of these diagrams 
contain, respectively, the position, the energy and the
$(-\infty,\infty)$-``energy'' representations. For anti-resonances, the 
diagrams are analogous.

\section{The time evolution of the Gamow states}
\label{sec:semievolut}

We are now going to obtain the time evolution of the Gamow states by 
extending the time evolution operator $\rme ^{-\rmi Ht/\hbar}$ into 
the spaces ${\mathbf \Phi}_{\rm exp}^{\prime}$ and 
${\mathbf \Phi}_{\rm exp}^{\times}$. Since such extension was constructed
in~\cite{LS2} to obtain the time evolution of the analytically continued 
Lippmann-Schwinger bras and kets, and since the Gamow states can be obtained
from the analytically continued Lippmann-Schwinger bras and kets, the
time evolution of the Gamow states will easily follow from the results 
of~\cite{LS2}. 

Let us calculate first the time evolution of the Gamow
bra $\langle ^{+}z_n|$ for a resonance energy:
\begin{eqnarray}
       \langle ^{+}z_n| \rme ^{-\rmi Ht/\hbar}|\varphi ^{+}\rangle &=&
       \langle ^{+}\widehat{z}_n| \rme ^{-\rmi \widehat{H}t/\hbar}|
           \widehat{\varphi}^{+}\rangle \nonumber \\
      &=& \langle ^{+}\widehat{z}_n| \rme ^{\rmi \widehat{H}t/\hbar}
           \widehat{\varphi}^{+}\rangle \, , 
              \quad t<0 \ {\rm only}\nonumber \\
       &=& \rme ^{\rmi z_nt/\hbar}
           \widehat{\varphi}^{+}(z_n) \, ,  \quad t<0 \ {\rm only}\nonumber \\
       & =& \rme ^{\rmi z_nt/\hbar} 
           \langle ^{+}z_n| \varphi ^{+}\rangle
               \, , \quad t<0 \ {\rm only}\, , \quad 
           \forall \widehat{\varphi}^+ \in \widehat{\mathbf \Phi}_{+{\rm exp}}  
                   \, ;
          \label{proofbraplustime}
\end{eqnarray}
that is,
\begin{equation}
      \langle ^+z_n| \rme ^{-\rmi Ht/\hbar}= 
         \rme ^{\rmi z_n t/\hbar} \langle ^+z_n| \, ,
        \quad {\rm only\ for}\ t<0  \, , \ n=1,2, \ldots \, .
      \label{adjoiotevbra1ares+}
\end{equation}
The reason why the time evolution for a Gamow bra associated with a 
resonant energy $z_n$ is defined only for $t<0$ is that when $t>0$
and $z_n=E_n-\rmi \Gamma _n/2$, the
factor $\rme ^{\rmi z_nt/\hbar}$ blows up exponentially, and therefore
$\rme ^{\rmi \widehat{H}t/\hbar} \widehat{\varphi}^{+}$ violates the
bound~(\ref{blowup2}) below. Hence, 
$\rme ^{\rmi \widehat{H}t/\hbar} \widehat{\varphi}^+$ is not in 
$\widehat{\mathbf \Phi}_{\rm +exp}$ when $t>0$, and therefore the dual 
extension of $\rme ^{\rmi \widehat{H}t/\hbar}$ is not well 
defined~\cite{EXPLA1}. We should also note that, strictly speaking, 
Eq.~(\ref{proofbraplustime}) does not prove that the time evolution
of $\langle ^+z_n|$ is well defined for $t<0$ in the sense of the theory
of distributions. In order to prove so, one needs that
the space of test functions ${\mathbf \Phi}_{\rm exp}$ be invariant under 
the action of $\rme ^{\rmi Ht/\hbar}$ for $t<0$. Since it is not known whether
${\mathbf \Phi}_{\rm exp}$ is invariant under the action of 
$\rme ^{\rmi Ht/\hbar}$, it remains an open problem to show that 
Eq.~(\ref{adjoiotevbra1ares+}) holds in the sense of the theory of
distributions.

Let us now calculate the time evolution of the Gamow
bra $\langle ^{+}z_n|$ for an anti-resonance energy:
\begin{eqnarray}
       \langle ^{+}z_n| \rme ^{-\rmi Ht/\hbar}|\varphi ^{+}\rangle &=&
       \langle ^{+}\widehat{z}_n| \rme ^{-\rmi \widehat{H}t/\hbar}|
           \widehat{\varphi}^{+}\rangle \nonumber \\
      &=& \langle ^{+}\widehat{z}_n| \rme ^{\rmi \widehat{H}t/\hbar}
           \widehat{\varphi}^{\pm}\rangle \, , 
              \quad t>0 \ {\rm only}\nonumber \\
       &=& \rme ^{\rmi z_nt/\hbar}
           \widehat{\varphi}^{+}(z_n) \, ,  \quad t>0 \ {\rm only}\nonumber \\
       & =& \rme ^{\rmi z_nt/\hbar} 
           \langle ^{+}z_n| \varphi ^{+}\rangle
               \, , \quad  t>0 \ {\rm only} \, ,  \quad 
           \forall \widehat{\varphi}^+ \in \widehat{\mathbf \Phi}_{+{\rm exp}}  
                   \, ;
          \label{proofbraplustimean}
\end{eqnarray}
that is,
\begin{equation}
      \langle ^+z_n| \rme ^{-\rmi Ht/\hbar}= 
         \rme ^{\rmi z_n t/\hbar} \langle ^+z_n| \, ,
        \quad {\rm only\ for}\ t>0  \, , \ n=-1,-2, \ldots \, . 
      \label{adjoiotevbra1ares+an}
\end{equation}
The reason why the time evolution for a Gamow bra associated with an 
anti-resonant energy $z_n=E_n+\rmi \Gamma _n/2$ is defined only 
for $t>0$ is that when $t<0$, the
factor $\rme ^{\rmi z_nt/\hbar}$ blows up exponentially, and therefore
$\rme ^{\rmi \widehat{H}t/\hbar} \widehat{\varphi}^+$ violates the
bound~(\ref{blowup2}). Hence, 
$\rme ^{\rmi Ht/\hbar} \widehat{\varphi}^+$ is not in 
$\widehat{\mathbf \Phi}_{\rm +exp}$ when $t<0$, and therefore the dual 
extension of $\rme ^{\rmi \widehat{H}t/\hbar}$ is not well 
defined~\cite{EXPLA1}. As in the case of Eq.~(\ref{proofbraplustime}),
Eq.~(\ref{proofbraplustimean}) does not prove that the time evolution
of $\langle ^+z_n|$ is well defined for $t>0$ in the sense of the theory
of distributions when $z_n$ is an anti-resonance energy.

The time evolution of the Gamow ket $|z_n^{-} \rangle$ associated with
a resonant energy $z_n$ is given by 
\begin{eqnarray}
        \langle \varphi ^{-}| \rme ^{-\rmi Ht/\hbar}|z_n^{-}\rangle &=&
        \langle \widehat{\varphi}^{-}| \rme ^{-\rmi \widehat{H}t/\hbar}|
             \widehat{z}_n^{-}\rangle  \nonumber \\
         &=& \langle \rme ^{\rmi \widehat{H}t/\hbar}\widehat{\varphi}^{-}|
             \widehat{z}_n^{-}\rangle  \nonumber\\
         &=& \left( \rme ^{\rmi z_n^*t/\hbar} \widehat{\varphi}^{-}(z_n^*) \right) ^*
           \, ,  \quad t>0 \ {\rm only}   \nonumber \\
          &=& \rme ^{-\rmi z_nt/\hbar} \left( \widehat{\varphi}^{-}(z_n^*) \right)^*
            \, ,  \quad t>0 \ {\rm only}      \nonumber \\
          &=& \rme ^{-\rmi z_n t/\hbar} \langle \varphi ^{-}|z_n^{-}\rangle 
       \, ,  \quad t>0 \ {\rm only}\, ,  \qquad 
           \forall \widehat{\varphi}^+ \in \widehat{\mathbf \Phi}_{+{\rm exp}} 
                   \, ;
\end{eqnarray}
that is,
\begin{equation}
      \rme ^{-\rmi Ht/\hbar}|z_n^{-}\rangle =
          \rme ^{-\rmi z_n t/\hbar} |z_n^{-}\rangle \, , \quad
          {\rm only\ for} \ t>0 \, , \ 
            n=1,2, \ldots \, . 
      \label{adjoiotev2a22}
\end{equation}
Similarly to Eqs.~(\ref{adjoiotevbra1ares+}) and (\ref{adjoiotevbra1ares+an}),
Eq.~(\ref{adjoiotev2a22}) is clearly not defined for $t<0$, although it 
remains to be proved that it holds for $t>0$ in a distributional way.

When we consider an anti-resonance, it can be
easily shown that Eq.~(\ref{adjoiotev2a22}) becomes
\begin{equation}
      \rme ^{-\rmi Ht/\hbar}|z_n^{-}\rangle =
          \rme ^{-\rmi z_n t/\hbar} |z_n^{-}\rangle \, , \quad 
            {\rm only\ for} \ t<0 \, , \ 
           n=-1,-2, \ldots  \, .
      \label{adjoiotev2a22an}
\end{equation}
Similarly to Eqs.~(\ref{adjoiotevbra1ares+}), (\ref{adjoiotevbra1ares+an})
and (\ref{adjoiotev2a22}), Eq.~(\ref{adjoiotev2a22an}) is clearly not 
defined for $t>0$, although it remains to be proved that it holds for 
$t<0$ in a distributional way.

In summary, the time evolution of the Gamow states is given by non-unitary 
semigroups and therefore is time asymmetric, expressing 
the irreversibility of a decaying process. Such semigroups
are simply (retarded or advanced) propagators that incorporate 
causal boundary conditions through the analytical properties of the
test functions~\cite{LS2}. However, as explained
above, the rigorous proof of (\ref{adjoiotevbra1ares+}), 
(\ref{adjoiotevbra1ares+an}), (\ref{adjoiotev2a22}) and (\ref{adjoiotev2a22an})
is still lacking, because it is not known whether 
${\mathbf \Phi}_{\rm exp}$ is invariant under $\rme ^{-\rmi Ht/\hbar}$.

\section{Resonance expansions}
\label{sec:resexp}

The Lippmann-Schwinger bras and kets are basis vectors that were used to 
expand normalizable, smooth wave functions in~\cite{LS1}:
\begin{equation}
     \langle r|\varphi ^{\pm} \rangle = 
        \int_0^{\infty}\rmd E \ \langle r|E^{\pm} \rangle 
                \langle ^{\pm}E|\varphi ^{\pm} \rangle \, .
         \label{lscomplerelat}
\end{equation}
The Gamow states are also basis vectors. The expansion generated by the Gamow 
states is called the resonance expansion. 


A given quantity (wave function, amplitude, etc.) can be expanded 
by resonance states in many different ways, depending on how many resonances 
we include in the expansion, see e.g.~review~\cite{05CJP}. When we include 
only a few
resonances close to the real axis, as in Berggren's and Berggren-like 
resonance expansions, the wave functions $\varphi (r)$ must fall off at 
infinity faster than exponentials~\cite{BOLLINI1,BOLLINI2}. However,
when we include all the resonances, we will
see that the wave functions must fall off faster than Gaussians.

\begin{figure}[ht]
\hskip-0.5cm \includegraphics{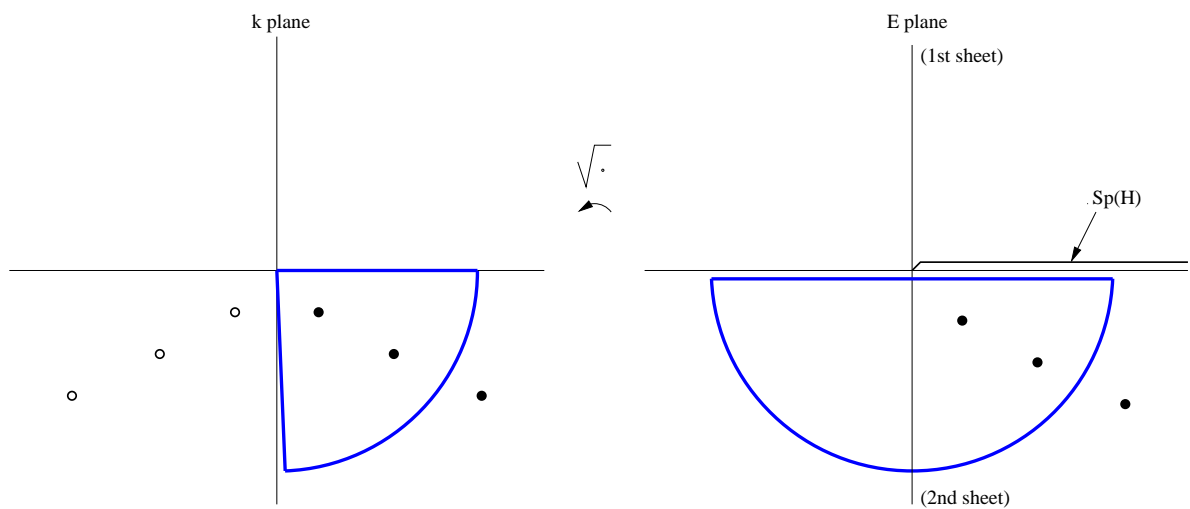}
\caption{The contour to obtain resonance expansions in the $k$-plane (left)
and in the $E$-plane (right). It is assumed that
the contour encloses all the resonances in the lower half plane of the
second sheet, and that the radius of the arc is sent to infinity. The filled
(hollow) dots represent the resonance (anti-resonance) poles.}
\label{fig:contour}
\end{figure}

For the sake of simplicity, we will focus on the resonance expansion 
of the transition 
amplitude from an ``in'' state $\varphi ^+$ into an ``out'' state 
$\varphi ^-$:
\begin{equation}
        \left( \varphi ^-,\varphi ^+\right)= 
         \int_0^{\infty} \rmd E \, \langle \varphi ^-|E^-\rangle S(E)
        \langle ^+E|\varphi ^+\rangle \, ,
        \label{superequation}
\end{equation}
where $S(E)$ is the $S$ matrix. For the spherical shell potential,
and also for any spherically symmetric potential that falls off
faster than exponentials, the $S$-matrix and the Lippmann-Schwinger
eigenfunctions can be analytically continued to the whole complex plane
(see Appendix~A of Ref.~\cite{NPA08}, and references therein). Thus,
by using the contour of Fig.~\ref{fig:contour}, we obtain
\begin{equation}
     \hskip-1cm
      \left( \varphi ^-,\varphi ^+\right)
      =  \sum_{n=1}^{\infty}
      \langle \varphi ^-|z_n^-\rangle\langle ^+z_n|\varphi ^+\rangle +
    \int_0^{-\infty} \rmd E \, \langle \varphi ^-|E^-\rangle S(E) 
      \langle ^+E|\varphi ^+\rangle  \, ,
      \label{zigzag}
\end{equation}
where we have tacitly assumed that
$\langle \varphi ^-|E^-\rangle S(E) \langle ^+E|\varphi ^+\rangle$ tends to 
zero in the infinite arc of the lower half plane of the second sheet. The 
integral in Eq.~(\ref{zigzag}) is done infinitesimally 
below the negative real semiaxis of the second sheet. By omitting
$\varphi ^-$ in~(\ref{zigzag}), we obtain the resonance expansion 
of the ``in'' wave functions, 
\begin{equation}
      \varphi ^+ = \sum_{n=1}^{\infty}
	   |z_n^- \rangle \langle ^+z_n|\varphi ^+\rangle +
    \int_0^{-\infty} \rmd E \, |E^-\rangle S(E) 
	\langle ^+E|\varphi ^+\rangle  \, .
        \label{states}
\end{equation}
The resonance expansion for the ``out'' wave function $\varphi ^-$ can be 
obtained in a similar way. In Eqs.~(\ref{zigzag}) and (\ref{states}), the 
infinite sum exhibits explicitly the contribution from the resonances, 
while the integral is the non-resonant background. 

In obtaining Eqs.~(\ref{zigzag}) and (\ref{states}), we have 
tacitly assumed that 
$\langle \varphi ^-|E^-\rangle S(E) \langle ^+E|\varphi ^+\rangle$ tends to 
zero in the infinite arc of the lower half plane of the second 
sheet. However, as shown in~\cite{LS2}, 
$\langle \varphi ^-|E^-\rangle S(E) \langle ^+E|\varphi ^+\rangle$ 
diverges exponentially there, since for any 
$\beta >0$ there is a constant $C$ such that~\cite{LS2}
\begin{equation}
     \left|  \langle \varphi ^-|z^-\rangle  \right| \leq C |q|^{-1/2} 
       \rme^{\frac{\, |{\rm Im}(q)|^2 \, }{2\beta}}  \, ,
           \label{blowup1} 
\end{equation}
\begin{equation}
     \left|  \langle ^+z|\varphi ^+\rangle \right| \leq C |q|^{-1/2} 
       \rme^{\frac{\, |{\rm Im}(q)|^2 \, }{2\beta}}  \, , 
           \label{blowup2} 
\end{equation}
where $q$ is the corresponding complex wave number in the fourth quadrant of
the $k$-plane. Therefore, Eqs.~(\ref{zigzag}) and (\ref{states}) 
need to be established properly. In order to do so, one has to control
the exponential blowups~(\ref{blowup1}) and (\ref{blowup2}) by 
calculating the time evolution of Eqs.~(\ref{zigzag}) and (\ref{states}):
\begin{eqnarray}
          \hskip-0.6cm
      &&\left( \varphi ^-, \rme ^{-\rmi Ht/\hbar} \varphi ^+\right)
      = \sum_{n=1}^{\infty} \rme ^{-\rmi z_nt/\hbar}
      \langle \varphi ^-|z_n^-\rangle\langle ^+z_n|\varphi ^+\rangle +
    \int_0^{-\infty} \rmd E \,
       \rme ^{-\rmi Et/\hbar} \langle \varphi ^-|E^-\rangle S(E) 
      \langle ^+E|\varphi ^+\rangle  , \nonumber   \\
      && \quad  \hskip12cm t>0 \ {\rm only,} 
      \label{alphazigzagtime}
\end{eqnarray}
\begin{eqnarray}
        \hskip-1cm
      \rme ^{-\rmi Ht/\hbar}\varphi ^+ = \sum_{n=1}^{\infty}
             \rme ^{-\rmi z_nt/\hbar}
	   |z_n^- \rangle \langle ^+z_n|\varphi ^+\rangle +
    \int_0^{-\infty} \rmd E \,
       \rme ^{-\rmi Et/\hbar}  |E^-\rangle S(E) 
	\langle ^+E|\varphi ^+\rangle  
        \, ,  \nonumber  \\
     \hskip10cm t>0 \ {\rm only.} 
        \label{alphastatestime}
\end{eqnarray}
These equations are valid because the following limits hold in the
infinite arc of the lower half plane of the second sheet for any 
$\alpha >0$:
\begin{equation}
     \lim_{z\to \infty} \rme ^{-\rmi \alpha z}\langle \varphi ^-|z^-\rangle = 
     \lim_{z\to \infty} \rme ^{-\rmi \alpha z}\langle ^+z|\varphi ^+\rangle 
          =0 \, ,
         \label{infinarli}
\end{equation}
which in turn follow from Eqs.~(\ref{blowup1}) and 
(\ref{blowup2}) (see however~\cite{EXPLA1}). Equations~(\ref{zigzag}) 
and~(\ref{states}) should then be understood as the limit of 
Eqs.~(\ref{alphazigzagtime}) and (\ref{alphastatestime}) when $t \to 0^+$. 





As shown in~\cite{BOLLINI1,BOLLINI2}, the Gamow bras and kets are already
well defined when the tails of the test functions fall off like exponentials
rather than like Gaussians. The reason why we chose a Gaussian falloff has
finally become clear. For test functions with exponential falloff, the above 
resonance expansions make no sense, since there is no way we can regularize 
the blowup of such test functions in the infinite arc of the second 
sheet~\cite{ROCCA}. However, imposing
a Gaussian falloff on the elements of ${\mathbf \Phi}_{\rm exp}$ enables
us to regularize their blowup in the complex energy plane by 
using the time evolution phase $\rme ^{-\rmi zt/\hbar}$ as a regulator. Also,
it is clear that Gaussian falloff is the slowest falloff that 
can be regularized in this way. 


As is well known, resonance expansions allow us to understand the 
deviations from exponential 
decay. If a particular resonance, say resonance 1, is dominant, then 
Eq.~(\ref{alphastatestime}) can be written as
\begin{eqnarray}
        \hskip-1cm
      \rme ^{-\rmi Ht/\hbar}\varphi ^+ = \rme ^{-\rmi z_1t/\hbar}
	   |z_1^- \rangle \langle ^+z_1|\varphi ^+\rangle +
           {\rm background}(1) \, ,
        \label{alphastatestime1}
\end{eqnarray}
where the term ``${\rm background}(1)$'' carries the contributions not
associated with resonance 1, including those from other resonances. Because
``${\rm background}(1)$'' never vanishes, there are always
deviations from exponential decay. The exponential law holds only when
the wave function is well tuned around the Gamow state $|z_1^- \rangle$, 
in which case ``${\rm background}(1)$'' can be neglected and only the 
resonance (Gamow state) contribution to the probability needs to be taken 
into account.


\section{Physical meaning of the Gamow states}
\label{sec:phmean}

We are now going to explore the physical meaning of the Gamow states. We will
do so by way of two analogies. The first analogy is that between
classical Fourier expansions, quantum completeness relations and
resonance expansions. The second analogy is that between the Gamow 
states and the quasinormal modes of classical systems. As always when
one draws analogies between classical and quantum mechanics, one should keep 
in mind that in classical mechanics the solutions of the wave equations 
are actual waves, whereas in quantum mechanics the solutions of the 
Schr\"odinger equation are probability amplitudes.

\subsection{Plane waves, the Lippmann-Schwinger bras and kets, and the
Gamow states}

Plane waves $\rme^{\rmi kx}$ represent monochromatic light 
pulses of well-defined wave number $k$. Experimentally,
one cannot prepare monochromatic plane waves: all that one can prepare
are wave packets $\widehat{\varphi}(k)$ that have some wave-number spread. The 
corresponding wave packet in the position representation, $\varphi (x)$, can be 
expanded in terms of the plane waves as
\begin{equation}
  \varphi (x) = \frac{1}{\sqrt{2\pi }} \int \rmd k \, 
                    \rme^{\rmi kx}\widehat{\varphi}(k) \, ,
       \label{fourinnet}
\end{equation}
which in Dirac's notation is written as
\begin{equation}
      \langle x|\varphi \rangle = \int \rmd k \, 
     \langle x|k \rangle \langle k|\varphi \rangle   \, .
\end{equation}
When $\widehat{\varphi}(k)$ is highly peaked around
a particular wave number $k_0$, the wave packet is well 
approximated by a monochromatic plane wave, 
$\varphi (x)\sim \rme ^{\rmi k_0x}$. 

The Lippmann-Schwinger bras and kets are a quantum version of
the classical plane waves. The monoenergetic eigenfunctions 
$\langle r|E^{\pm}\rangle$ represent a particle with a sharply defined 
energy $E$ (and with additional ``in'' or ``out'' boundary conditions). In 
analogy to the Fourier
expansion of wave packets in terms of classical plane waves,
Eq.~(\ref{fourinnet}), the eigenfunctions $\langle r|E^{\pm}\rangle$
expand wave functions $\varphi ^{\pm}$ as in Eq.~(\ref{lscomplerelat}). When 
the wave packet $\widehat{\varphi}^{\pm}(E)$ is highly peaked around a 
particular energy $E_0$, then the approximation 
$\varphi ^{\pm}(r) \sim \langle r|E_0^{\pm}\rangle$ holds.

The physical meaning of the Gamow states is similar. Likewise 
the monoenergetic scattering states, the Gamow states cannot
be prepared experimentally: All that can be prepared is a wave packet
$\varphi ^+$. In complete analogy to the expansions~(\ref{fourinnet})
and (\ref{lscomplerelat}), the Gamow states and an additional set of
``background'' states expand a wave function $\varphi ^+$, see
Eq.~(\ref{states}). When the wave function is finely tuned around 
one resonance, say resonance 1, then in general the approximation 
$\varphi ^{+}(r) \sim \langle r|z_1\rangle$ holds for all practical 
purposes~\cite{SIGS}. It is in this sense that a lone Gamow 
state is the wave function of a quantum decaying particle. 

When the approximation $\varphi ^{+}(r) \sim \langle r|z_1\rangle$ holds,
the Gamow state can be used to
characterize the transport of probability
in a time-dependent description of resonant scattering and decay. For instance,
in Ref.~\cite{GASTON07} Garcia-Calderon {\it et al.}~present the example 
of a delta-shell potential where one resonance dominates the decay of the 
system. In order to show so, the authors
of~\cite{GASTON07} calculate the survival probability using a square
integrable function $\varphi ^+$. They also use a resonant
expansion to approximate $\varphi ^+$ by one single resonant state
$\langle r|z_1\rangle$. As shown in Fig.~3 of Ref.~\cite{GASTON07}, the 
exponential decay of the survival
probability calculated by way of the Gamow state $\langle r|z_1\rangle$ is
indistinguishable from the one calculated by way of the ``exact'' 
square integrable wave function $\varphi ^+$.


\subsection{Quasinormal modes vs.~resonance states}

In classical mechanics, confined linear oscillating systems --e.g., 
finite strings, membranes or cavities filled with electromagnetic
radiation-- have preferred states of motion. Such states of motion are 
called normal modes. Each normal mode is associated with a characteristic
real frequency. Unless it is perturbed, a system in a normal mode will keep
vibrating the same way perpetually. When friction or dissipation enters into 
play and therefore the system dissipates energy, the system has 
preferred ways of doing so, which are called the quasinormal modes. Unconfined 
linear oscillating systems also have quasinormal modes, and they
are obtained by imposing Sommerfeld's radiation condition, which is the
classical counterpart of the POBC~\cite{NESTERENKO}. Each
quasinormal mode is associated with a characteristic complex frequency, whose
imaginary part is associated with the exponential damping of the oscillation.

In quantum mechanics, normal modes correspond to bound states, and 
quasinormal modes correspond to resonance states. Much like quasinormal
modes describe the system's preferred ways of dissipating energy, the
Gamow states describe the system's preferred ways of decaying. The imaginary 
part of the complex energy of the Gamow state is associated with exponential 
decay, in analogy to the imaginary part of the complex, classical frequency
being associated with exponential dissipation.

\subsection{Physical meaning of the exponential blowup of the Gamow states}

The Gamow states blow up exponentially at infinity, and it is important
to understand the physical origin of such exponential blowup. Let us consider
first the Lippmann-Schwinger eigenfunction $\langle r|E^+\rangle$. These
time-independent, non-normalizable eigenfunctions are interpreted as an 
incoming plane 
wave that impinges on a target and an outgoing wave multiplied by the 
$S$ matrix. However, the actual expression of $\langle r|E^+\rangle$ does 
not lead to such interpretation. Only when one views the Lippmann-Schwinger
eigenfunction in a time-dependent fashion, one can arrive at such 
interpretation. Thus, even though they are time-independent, the 
Lippmann-Schwinger eigenfunctions encode what happens in a scattering 
experiment at all times.

Similarly, the Gamow states are time-independent, 
non-normalizable eigenfunctions that describe the decay
of a quantum system at all times. Since after a long time (formally,
when $t\to \infty$) the resonance will surely have decayed and gone
to infinity, the Gamow state needs to provide a time-independent probability 
amplitude of finding the particle at infinity that
is much greater than the probability of finding the particle anywhere else in
space, hence the exponential blowup at infinity.

\section{Conclusions}
\label{sec:conclusions}

We have used the spherical shell potential to present a systematic 
procedure to construct the rigged Hilbert space of the Gamow states. A 
Gamow state has been defined as the solution of the homogeneous integral 
equation introduced in~\cite{WOLF,MONDRAGON84}. Such integral equation is 
of the Lippmann-Schwinger type, and is equivalent to the Schr\"odinger equation 
subject to the POBC. 

By applying the theory of distributions, we have constructed the Gamow bras 
and kets and shown that they are, respectively, linear and antilinear
functionals over the space of test functions ${\mathbf \Phi}_{\rm exp}$,
where the elements of ${\mathbf \Phi}_{\rm exp}$ are smooth
functions that fall off faster than Gaussians. We have shown that 
the Gamow bras and kets are, respectively, ``left'' and ``right'' eigenvectors
of the Hamiltonian, and that their associated eigenvalues 
coincide with the resonance energies. We have argued, although not rigorously
proved, that the exponential time evolution is given by a non-unitary 
semigroup. Such semigroup time evolution exhibits the time asymmetry 
of a decaying process.

We have also constructed the energy representations of the Gamow states. We
have shown that such energy representations are given by either the
complex delta function or by the residue distribution. These results complement
the properties of the Gamow states in the momentum representation
obtained in~\cite{MONDRAGON84,BOLLINI1,BOLLINI2}.

Because in the position representation the wave functions in 
${\mathbf \Phi}_{\rm exp}$ fall off faster than Gaussians, we have been able
to construct resonance expansions that include all the resonances. Such 
resonance expansions also exhibit the time asymmetry of the decaying 
process.

Finally, we have clarified some of the physical properties of the Gamow
states by drawing analogies with classical Fourier expansions and quasinormal
modes. We have also clarified the origin of the exponential blowup of 
a Gamow state at infinity.



\section{Acknowledgment}
\label{sec:ack}

I am indebted to Prof.~Alfonso Mondrag\'on for his careful and patient 
explanations on Eq.~(\ref{Monlisus}). This research has been partially 
supported by Ministerio de Ciencia e Innovaci\'on of Spain under 
project TEC2011-24492.


{

\appendix

\section{Proofs}
\label{sec:proofsofprop}

Here we list the proofs of some results we stated in the paper. In
the proofs, whenever an operator $A$ is acting on the bras, we will use
the notation $A^{\prime}$, and whenever it is acting on the kets, we will
use the notation $A^{\times}$:
\begin{equation}
    \langle ^{\pm}z_n|A^{\prime}|\varphi ^{\pm} \rangle := 
    \langle ^{\pm}z_n|A^{\dagger} \varphi ^{\pm} \rangle  \, , 
         \qquad  \forall  \varphi ^{\pm} \in {\mathbf \Phi}_{\rm exp} \, ,
         \label{beigenrhsp}
\end{equation}
\begin{equation}
    \langle \varphi ^{\pm}|A^{\times}|z_n^{\pm}\rangle :=
    \langle A^{\dagger}\varphi ^{\pm}|z_n^{\pm}\rangle \, , 
           \qquad   \forall \varphi ^{\pm}\in {\mathbf \Phi}_{\rm exp} \, . 
       \label{keigenrhsc}
\end{equation}
Thus, $A^{\prime}$ denotes the \emph{dual} extension of $A$ acting to the 
left on the elements of ${\mathbf \Phi}_{\rm exp}^{\prime}$, whereas
$A^{\times}$ denotes the \emph{antidual} extension of $A$ acting to 
the right on the elements of ${\mathbf \Phi}_{\rm exp}^{\times}$. This 
notation stresses that $A$ is acting outside the Hilbert space and specifies 
toward what direction the operator is acting, thereby making the proofs more 
transparent.

\vskip0.5cm

\begin{proof}[{\bf Proof of Proposition~2}] \quad 

The proofs of Eqs.~(\ref{-ErepGb})-(\ref{+ErepGk}) all follow the same 
pattern. We start by proving~(\ref{-ErepGk}). The Gamow eigenfunction 
$u(r;z_n)$ is proportional to the analytic 
continuation of the Lippmann-Schwinger eigenfunction
$\chi ^-(r;E)$~\cite{COCOYOC},
\begin{equation}
       u(r;z_n)= \rmi \sqrt{2\pi \, } {\cal N}_n \chi ^-(r;z_n) \, .
       \label{jaequai}
\end{equation}
From this equation and from the analytic properties of the elements
$\widehat{\varphi}^- \in {\mathbf \Phi}_{-{\rm exp}}$ obtained in~\cite{LS2},
it follows that
\begin{eqnarray}
      \langle \widehat{\varphi}^-|\widehat{z}_n^- \rangle &=&
       \langle \widehat{\varphi}^-|U_-^{\times}|z_n^- \rangle \nonumber \\
       &=& \langle U_-^{\dagger}\widehat{\varphi}^-|z_n^- \rangle \nonumber \\
       &=& \langle \varphi ^-|z_n^- \rangle \nonumber \\
       &=& \int_0^{\infty}\rmd r \, [\varphi ^-(r)]^* u(r;z_n) 
          \hskip1cm  \mbox{by~(\ref{Gketdef})}  \nonumber \\
       &=& \rmi \sqrt{2\pi \, } \,  {\cal N}_n  \int_0^{\infty}\rmd r \, 
            [\varphi ^-(r)]^* \chi ^-(r;z_n)
          \hskip1cm  \mbox{by~(\ref{jaequai})}  \nonumber \\
       &=& \rmi \sqrt{2\pi \, } \,  {\cal N}_n  \, 
           [\widehat{\varphi}^-(z_n^*)]^* \nonumber \\
       &=& \rmi \sqrt{2\pi \, } \,  {\cal N}_n 
           \langle \widehat{\varphi}^-|\widehat{\delta}_{z_n} \rangle 
                   \, , \quad
       \forall \widehat{\varphi}^- \in \widehat{\mathbf \Phi}_{-{\rm exp}}  
                   \, , 
       \label{profosjd}
\end{eqnarray}
which proves~(\ref{-ErepGk}). The proof of~(\ref{+ErepGb}) is analogous.

In order to prove~(\ref{+ErepGk}), we need the following 
relation~\cite{COCOYOC}:
\begin{equation}
       u(r;z_n)= - \frac{\sqrt{2\pi \, }}{{\cal N}_n} \,  
                   {\rm res} \, [\chi ^+(r;z)]_{z=z_n}  \, .
       \label{jaequaires}
\end{equation}
Then,
\begin{eqnarray}
         \hskip-1cm
      \langle \widehat{\varphi}^+|\widehat{z}_n^+ \rangle &=&
       \langle \widehat{\varphi}^+|U_+^{\times}|z_n^+ \rangle \nonumber \\
       &=& \langle U_+^{\dagger}\widehat{\varphi}^+|z_n^+ \rangle \nonumber \\
       &=& \langle \varphi ^+|z_n^+ \rangle \nonumber \\
       &=& \int_0^{\infty}\rmd r \, [\varphi ^+(r)]^* u(r;z_n)
           \hskip1cm  \mbox{by~(\ref{Gketdef})}  
       \nonumber \\
       &=& - \frac{\sqrt{2\pi \,}}{{\cal N}_n}
         \int_0^{\infty}\rmd r \, 
            [\varphi ^+(r)]^*  {\rm res}\, [\chi ^+(r;z)]_{z=z_n}
         \hskip1cm \mbox{by~(\ref{jaequaires})}  \nonumber \\
       &=& - \frac{\sqrt{2\pi \,}}{{\cal N}_n} \,
            {\rm res} \, [\widehat{\varphi}^+(z_n^*)]^* \nonumber \\
       &=& - \frac{\sqrt{2\pi \,}}{{\cal N}_n}
            \langle \widehat{\varphi}^+| \widehat{\rm res}_{z_n} \rangle 
                   \, , \quad
       \forall \widehat{\varphi}^+ \in \widehat{\mathbf \Phi}_{+{\rm exp}}  
                   \, , 
       \label{profosjdres}
\end{eqnarray}
which proves~(\ref{+ErepGk}). The proof of~(\ref{-ErepGb}) is analogous.
\renewcommand{\qedsymbol}{}
\end{proof}

\vskip0.3cm

\begin{proof}[{\bf Proof of Eq.~(\ref{bwcomfrel})}] \quad 

Let $\widehat{\varphi}^-\in \widehat{\mathbf \Phi}_{-{\rm exp}}$. It was 
proved in~\cite{LS2} that for any $\beta >0$, the following 
estimate is valid in the lower half plane of the second sheet:
\begin{equation}
     \left| [\widehat{\varphi}^-(z^*)]^* \right| \leq C |q|^{-1/2} 
       \rme^{\frac{\, |{\rm Im}(q)|^2 \, }{2\beta}}  \, , 
\end{equation}
where $q$ is the corresponding complex wave number in the fourth quadrant
of the $k$-plane. This estimate 
implies that in the infinite arc of the lower half plane of the second sheet,
the following limit holds for any $\alpha >0$ (see however~\cite{EXPLA1}):
\begin{equation}
       \lim_{|z| \to \infty} \rme^{-\rmi \alpha z} [\widehat{\varphi}^-(z^*)]^*
          = 0 \, .
\end{equation}
Then, by Cauchy's formula,
\begin{equation}
     \rme^{-\rmi \alpha z_n} [\widehat{\varphi}^-(z_n^*)]^*= 
      -\frac{1}{2\pi \rmi} \int_{-\infty}^{\infty}\rmd E\,\rme^{-\rmi \alpha E}
      \frac{1}{E-z_n} \, [\widehat{\varphi}{^-}(E)]^* \, .
\end{equation}
Multiplying this equation by $\rmi \sqrt{2\pi} {\cal N}_n$ yields
\begin{equation}
          \hskip-0.5cm
      \rme^{-\rmi \alpha z_n}
      \rmi \sqrt{2\pi} {\cal N}_n[\widehat{\varphi}{^-}(z_n^*)]^*= 
         \int_{-\infty}^{\infty}\rmd E\,\rme^{-\rmi \alpha E}
        \left( -\frac{{\cal N}_n}{\sqrt{2\pi}} \right) 
      \frac{1}{E-z_n} \, [\widehat{\varphi}{^-}(E)]^* \, .
           \label{nexopso}
\end{equation}
From Eqs.~(\ref{samefunction}), (\ref{BWfunction}) and (\ref{nexopso}) it 
follows that
\begin{equation}
     \rmi \sqrt{2\pi} {\cal N}_n[\widetilde{\varphi}{^-}(z_n^*)]^*=
     \langle \widetilde{\varphi}^-|\frac{1}{E-z_n}{^-}\rangle \, .
      \label{nonumvsr}
\end{equation}
We now define the action of ${\mathbb A}^{\times}$ on 
$|\widehat{z}_n^-\rangle$ by
\begin{equation}
       \langle \widetilde{\varphi}^-|
        {\mathbb A}^{\times}|\widehat{z}_n^- \rangle :=
       \langle {\mathbb A}^{-1}\widetilde{\varphi}^-|\widehat{z}_n^-\rangle 
                       \, .
       \label{nonumvsrnex}
\end{equation}
Since
\begin{equation}
     \langle {\mathbb A}^{-1}\widetilde{\varphi}^-|\widehat{z}_n^-\rangle = 
     \langle \widehat{\varphi}^-|\widehat{z}_n^-\rangle =
     \rmi \sqrt{2\pi} {\cal N}_n[\widehat{\varphi}{^-}(z_n^*)]^*=
      \rmi \sqrt{2\pi} {\cal N}_n[\widetilde{\varphi}{^-}(z_n^*)]^* \, ,
       \label{nonumvsrnex2}
\end{equation}
we have that
\begin{equation}
       \langle \widetilde{\varphi}^-|
        {\mathbb A}^{\times}|\widehat{z}_n^- \rangle =
      \langle \widetilde{\varphi}^-|\frac{1}{E-z_n}{^-}\rangle \, ,
       \quad \forall \widetilde{\varphi}^- \in
        \widetilde{\mathbf \Phi}_{-{\rm exp}} \, ,
       \label{nonumvsrnex3}
\end{equation}
which proves (\ref{bwcomfrel}).

\renewcommand{\qedsymbol}{}
\end{proof}

}

\vskip3cm

\begin{thebibliography}{99}
\bibitem{CHC} Instead of ``complex Hermitian conjugation,'' one
can use the term ``Schwartz complex conjugation,'' since such operation
actually originates from the Schwarz reflection principle. This 
principle states
that for a holomorphic function in the upper complex plane, continuous
and real valued on the real line, one can write an analytic continuation
for the whole $\mathbb C$ plane such that $f(z^*)=f^*(z)$.
\bibitem{EXPLA1} Strictly speaking, we should have demanded that our 
test functions fall off at infinity in the position representation 
not like $\rme ^{-r ^{2}/2}$ but like $\rme ^{-\frac{r ^{2+\epsilon}}{2+\epsilon}}$, 
where $\epsilon$ is
any small positive number. This way, the exponential blowup of the test
functions in the energy representation would had been given by
$\rme^{\frac{\, |{\rm Im}(q)|^{2-\delta} \, }{(2-\delta)}}$, 
$\delta=\epsilon/(1+\epsilon) >0$ and small, 
rather than by Eqs.~(\ref{blowup1}) and (\ref{blowup2}). For the sake 
of clarity, we will ignore this technicality.
\bibitem{SIGS} Examples of square-integrable wave 
functions that approximate a Gamow state very closely can be found in 
Refs.~\cite{STRAUSS06,DURR11}.
\end{thebibliography}
\end{document}